\documentclass[aps,prd,reprint,a4paper,showpacs,nofootinbib,superscriptaddress]{revtex4-2}

\usepackage{hyperref}
\hypersetup{
  colorlinks=true,        
  linkcolor=blue,         
  citecolor=cyan,         
}

\usepackage{graphicx}
\usepackage{dcolumn}
\usepackage{bm}
\usepackage{color}
\usepackage{enumitem}
\usepackage{amsmath}

\usepackage{amssymb}

\newcommand{\beq}{\begin{equation}}
\newcommand{\eeq}{\end{equation}}
\newcommand{\bea}{\begin{eqnarray}}
\newcommand{\eea}{\end{eqnarray}}

\usepackage{bbm}
\usepackage{amsfonts}
\usepackage{mathrsfs}
\usepackage{latexsym}
\usepackage{epsfig}
\usepackage{epstopdf}
\usepackage{epstopdf}
\usepackage{graphicx}
\usepackage{amssymb}
\usepackage{amsmath}
\usepackage{dcolumn}
\usepackage{bm}
\usepackage{color}
\usepackage{comment}
\usepackage{xcolor}
\usepackage{dsfont}

\usepackage{amsmath}
\usepackage{lineno}
\usepackage{float, geometry,lipsum}
\usepackage{orcidlink}
\begin{document}

\title{\bf Harmonic oscillation  and  orbital morphology for spinning charged quantum corrected black hole}

\author{Rameez Khalid}
\email{rameezkhalid34@gmail.com}
\affiliation{Department of Mathematics, COMSATS University Islamabad, Sahiwal Campus-57000, Islamabad, Pakistan}

\author{Muhammad Yasir \orcidlink{0000-0003-0197-1725}}
\email{yasirciitsahiwal@gmail.com (Corresponding Author)}
\affiliation{Institute of Fundamental Physics and Quantum Technology, Department of Physics, School of Physical Science and Technology, Ningbo University, Ningbo, 315211, Zhejiang, People’s Republic of China}
\affiliation{Center for Theoretical Physics, Khazar University, 41 Mehseti Str., Baku, AZ1096, Azerbaijan}
\affiliation{Department of Mathematics, School of Science, University of Management and Technology,  Lahore, 54000, Pakistan}
\author{Shahid Qaisar}
\email{shahidqaisar90@ciitsahiwal.edu.pk}
\affiliation{Department of Mathematics, COMSATS University Islamabad, Sahiwal Campus-57000, Islamabad, Pakistan}

\author{Faisal Javed}
\email{faisaljaved.math@gmail.com (Corresponding Author)}
\affiliation{College of Transportation, Tongji University, Shanghai, 201804, China}

\begin{abstract}

This study investigates the dynamical behavior of test particles around a spinning charged quantum-corrected black hole (QCBH), \textcolor{black}{constructed within the $\gamma=0$ sector of the Einstein--Maxwell--dilaton (EMd) framework, by examining the effects of the spin parameter $a$, quantum correction parameter $b$, and electric charge parameter $Q$.} Using \textcolor{black}{the resulting QC spinning charged spacetime geometry}, we analyze the effective potential and effective force to determine the stability and structure of the circular orbits in a strong gravitational field. Furthermore, we calculate the radial $(\Omega_r)$, vertical $(\Omega_\theta)$, and angular $(\Omega_\phi)$ oscillation frequencies, as well as the corresponding Periapsis and Lense-Thirring precession frequencies. The spin parameter primarily determines the spinning properties of spacetime, while the quantum correction and \textcolor{black}{charge parameter introduce additional deviation from Kerr-Newman geometry, particularly near the event horizon}. These corrections alter the position and stability of the circular orbits, and the spectrum is effectively associated with the particle's motion. The results provide a comprehensive theoretical description of the orbital dynamics of a spinning \textcolor{black}{charged QCBH} may be through future high-precision astrophysical observations. 

\end{abstract}

\maketitle
\date{\today}

\section{Introduction}\label{introduction}
Black holes are one of the most remarkable predictions of Einstein's theory of general relativity, providing a systematic approach to analyzing the fundamental properties of gravity and high-energy astrophysical phenomena \cite{Einstein1915}. The BHs provide an excellent testing ground for probing strong gravitational effects, studying spacetime geometry, and improving our understanding of the interaction between matter and gravity \cite{Will2014, Abbott}. By the mid-twentieth century, the precise detection of BH solutions and the analysis of particle motion in curved spacetime had become an important research area, providing valuable insight into gravitational phenomena and the behavior of matter in astrophysical environments \cite{Schwarzchild1916, Riess1998}. With some theoretical challenges, GR has achieved remarkable success in describing gravitational interactions, but some theoretical and observational challenges still exist. Even so, it remains the fundamental framework for understanding compact objects and the large-scale dynamics of the universe \cite{Plebanski and Krasinski2024, d'Inverno1992, Hobson2006}. In recent years, researchers have devoted a great deal of effort to understanding how quantum corrections affect spinning charged BHs. To account for quantum gravitational effects, several studies have proposed QC Kerr BH models that build upon the classical Kerr solution that characterizes spinning charged BHs. These investigations suggest that quantum corrections can alter the observable BH properties and provide fresh insights into the behavior of  BHs \cite{C.Liu2020, M.Afrin2023}.
The scarcity of spinning charged QCBH models in quantum gravity makes observational verification of these theories extremely challenging. This lack of suitable models has prompted the use of the conformal BH metric, including models proposed by Lewandowski et al. \cite{Glampedakis, Zhang}. The spinning charged QCBH metric obtained through the algorithm of Newman–Janis with parameter $\alpha$, closely resembles the classical Kerr solution and gives a useful structure for evaluating quantum gravity theories by astrophysical examination \cite{Gong, Chen, Ali, Vachher}. Inspection of particle motion in the proximity of BHs is a central subject in relativistic astrophysics, as it offers an important understanding of the fundamental physical properties of spacetime. Therefore, considerable research has been carried out in this area \cite{Bardeen1972, Mashhoon1985, Stuchlik2010, Kolos2017, Oteev2018, Ren2020, Zahid2021}.

Previous studies have also analyzed classical BH models, such as Schwarzschild BH and Kerr BH, under both high-field and low-field conditions. However, substantial opportunities remain for investigating alternative theories of gravity, chiefly those involving cosmological relevance and dilaton scalar fields. Cosmological observations of smaller astronomical bodies have been utilized to test and evaluate these alternative gravitational models \cite{Bambi2017, Zhou2018}.
This work investigates a class of EMd gravity models in which the gravitational field, electromagnetic field, and scalar dilaton field are coupled together \cite{Kaluza1921, Klein1926, Ortin2015, Duff1986, Overduin1997}. In general, the aim is to analyze the physical and geometric behavior of these EMd models and explore their significance in gravitational and high-energy physics \cite{Gibbons1988, Garfinkle1991}. In this study, we investigate how quantum corrections alter the thermodynamic, dynamical, and observational properties of spinning \textcolor{black}{charged} QCBH \cite{Yasir2024Barrow,Yasir2024Thermal, Chaudhary2022Thermodynamic,Yasir2025Particle,Mushtaq2025Imprints}. \textcolor{black}{These corrections may produce theoretically distinguishable changes in particle motion, orbital dynamics, and related BH properties, but their observational implications require further investigation. }\cite{MYAJ0, MYAJ1, MYAJ3}.
\\\textcolor {black} {Recent studies have explored particle dynamics and rotational oscillations in the spacetime of various modified BHs. Mustafa et al. investigated particle motion and rotational frequencies around quantum modified BHs \cite{Mustafa2024}, while Ashraf et al. investigated similar phenomena in rotating Einstein-Yang-Mills-Higgs BHs \cite{Ashraf2025}. Zulqarnain et al. extended these analyses to rotating Bumblebee Kerr-Newman-AdS spacetime \cite{Zulqarnain2026}. In contrast, this study considers the spinning charged quantum correction geometry and examines the effects of spin, \textcolor{black}{electric} charge, and quantum correction on circular orbits, effective potential and force, rotational frequency, and relativistic precession. The innovation of this study lies in characterizing these combined effects and their deviation from the classical Kerr-Newman limit.}
This study explores the motion of test particles around a spinning \textcolor{black}{charged} and axially symmetric  QCBH. The main purpose is to examine how the BH parameters affect the trajectories and behavior of particles moving in the surrounding spacetime. In this framework, the QC geometry is characterized through three fundamental parameters that govern the physical properties of the solution.

The structure of this work is as follows: In Section II, we present the spacetime framework of spinning \textcolor{black}{charged} QCBHs and formulate the key equations that describe the particle motion in this geometry. In this section, the characteristics of circular trajectories, effective potentials, and the corresponding effective force on test particles are also examined. In Section III, we focus on harmonic oscillations around a stable circular motion. The radial, vertical, and angular frequencies $(\Omega_{r}~,~\Omega_{\theta}~,~\Omega_{\phi})$ are calculated, and the effects of periapsis precession are analyzed together with Lense–Thirring precession, and the results are compared with those of the classical Kerr-Newman BH. In Section IV, we discuss the comparison of results and highlight their significance for examining quantum gravity signatures through astrophysical observations of BH systems.

\section*{II.~~~THE EINSTEIN-MAXWELL-dilaton Theory.} 
We re-examine the fundamental formulation of Einstein's theory of gravity within the EMd framework. This theory extends the electrodynamics of charged by introducing quantum corrections that become significant in the presence of strong electromagnetic fields. It was initially proposed within the Kaluza-Klein theoretical framework \cite{Kaluza1921, Klein1926}. The EMd action is minimally coupled with charged electrodynamics, enabling a comprehensive description of the interaction between gravity and charged electromagnetic fields. This approach provides valuable insight into the spacetime behavior of strong-field regions where quantum charged electrodynamics corrections are not negligible. The resulting framework lays a crucial foundation for exploring the geometry, dynamics, and observational properties of charged compact objects, including BH, within a more comprehensive theoretical framework of gravity. We consider the single-parameter equation from EMd \cite{Herdeiro2025}
\begin{equation}
S=\frac{1}{4\pi}\int \left\{
\frac{1}{4}R\,\epsilon
-\frac{1}{2}d\Phi\wedge \star d\Phi
-\frac{f(\Phi)}{2}\mathcal{F}\wedge \star \mathcal{F}
\right\}.
\label{1}
\end{equation}
\begin{equation}
f(\Phi)=e^{-2\gamma\Phi}.
\label{2}
\end{equation}
where $R$ represents the Ricci scalar curvature, and $\epsilon$ represents spacetime volume. \textcolor{black}{We restrict our analysis to the $\gamma=0$ sector of the EMd theory. In this limit $f(\Phi)=1$, the scalar field can be consistently set to $\Phi=0$, thus simplifying the EMd action to the Einstein-Maxwell theory. Therefore, in the configuration considered in this paper, the dilaton field does not provide an independent dynamic contribution. The classical rotating charged configuration corresponds to Kerr-Newman geometry. Quantum effects are introduced by replacing the constant mass parameter $M$ with the radial mass function $M(r)$  defined in equation (\ref{4}). This substitution produces the QC rotating charged metric shown in equation (\ref{3}) while preserving the fundamental rotational and charged properties of spacetime \cite{Gibbons1988,Garfinkle1991}}.
 In this analysis, the spacetime geometry is represented by the Boyer-Lindquist coordinates $(t), (r), (\theta) ,(\phi)$, where the metric is determined by mass, \textcolor{black}{electric charge parameter, spin parameter, and quantum correction parameter.} These parameters collectively determine the spacetime geometry and physical characteristics of the BH. The corresponding line elements are given by the following equation: \cite{Herdeiro2025}  
\begin{eqnarray}\nonumber
 ds^2 &=&-\frac{(\Delta-a^2 sin^2 \theta)}{\Sigma} dt^2
 +\big(\frac{\Sigma}{\Delta}\big) dr^2+\Sigma d\theta^2\\ \nonumber
 &+&\big(\frac{(r^2+a^2)^2 -\Delta a^2sin^2\theta}{\Sigma}\big) sin^2 \theta d\phi^2 \\
&-&2asin^2 \theta \big(\frac{r^2+a^2-\Delta}{\Sigma}\big) dt d\phi,
\label{3}
\end{eqnarray}
where
\begin{eqnarray} \nonumber
\Delta&=&r^2-2M(r)r+a^2+Q^2,~~~M(r)=M-\frac{bM^4}{2r^3}, \\
\Sigma&=&r^2+a^2cos^2\theta.
\label{4}
\end{eqnarray}
The metric described in equation (\ref{3}) depends on three fundamental physical parameters. This formula derives different limiting cases. In particular, when $b=0$ and $Q\neq0$, the line element is a Kerr-Newman metric and if $b=0=Q$, then the line element is a Kerr metric. The event horizon is determined by solving the equation $\Delta(r)=0$, which can be expressed as:
\textcolor{black}{
\begin{equation}
r^2+a^2-2\big(M-\frac{bM^4}{2r^3}\big)r+Q^2=0. 
\label{5}
\end{equation}}
This \textcolor{black}{quartic} equation governs the positions of the \textcolor{black}{inner and outer event} horizons. In its solutions, the largest positive root represents the radius of the outer event horizon, while smaller positive roots \textcolor{black}{(when they exists)} represent the radius of the inner horizon. Introducing the parameter $b$ alters the radial position of the horizon relative to the Kerr-Newman geometry, thus modifying the horizon structure. \textcolor{black}{In the classical limit $b \to 0$, the quantum correction disappears and the standard Kerr-Newman horizon structure is restored}. These deviations become increasingly significant near the BH (the region of strongest gravitational effects). At a smaller radial distance, the effects of the quantum correction are expected to be even more pronounced. Therefore, the modified geometry differs significantly from conventional Kerr-Newman spacetime strong-field regions. These changes could affect the physical properties of the BH and the motion of nearby particles. Thus, the parameter $b$ provides an effective way to study quantum-inspired BH horizon correction. \textcolor{black}{These results give further insight into the horizon structure of the spinning \textcolor{black}{charged} QCBH spacetime.}  
\begin{figure}[H]
\centering
\includegraphics[width=85mm,height=60mm]{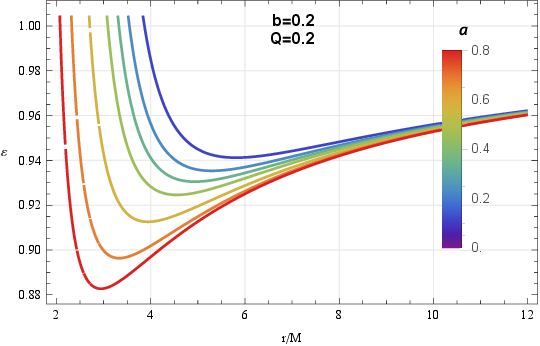}
\includegraphics[width=85mm,height=60mm]{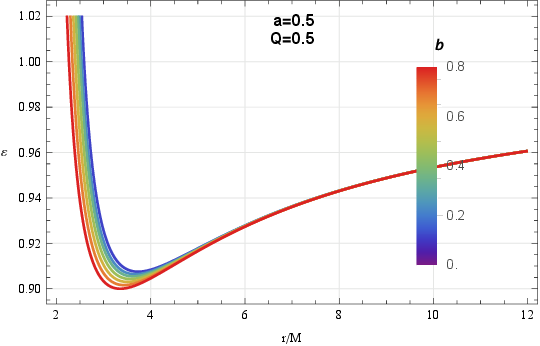}
\includegraphics[width=85mm,height=60mm]{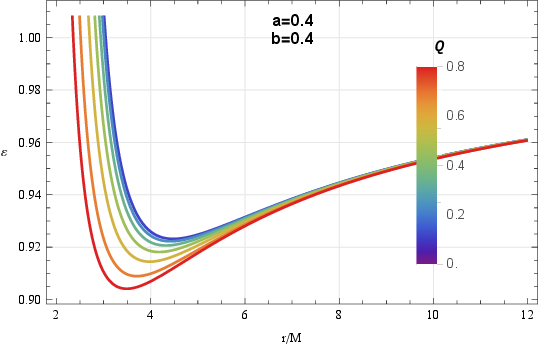}
\caption{Energy $\mathcal{E}$ evolution of test particles orbiting a \\ \phantom{FIG.1.}~~~spinning \textcolor{black}{charged} QCBH } 
\label{fig.1}
\end{figure}

\section*{A.\hspace{0.5cm}Circular orbits around spinning charged QCBH}
The motion of the particles near a spinning \textcolor{black}{charged} QCBH provides a valuable perspective for understanding the interaction between gravity, electromagnetic, and spinning spacetime geometry. In spinning \textcolor{black}{charged} QCBH, particle trajectories are influenced not only by the QCBH's mass and angular momentum but also by its charged quantum correction effects. The spinning of the \textcolor{black}{charged} QCBH creates a frame-dragging effect, causing nearby particles to experience spacetime distortion, thus significantly altering their orbital behavior. \textcolor{black}{For the neutral test particle considered in this study, the BH \textcolor{black}{electric} charge parameter alters the surrounding spacetime geometry. Therefore, its effect on the particle motion is not through direct electromagnetic interaction, but rather indirectly through these geometric changes, while} quantum corrections further alter the spacetime structure, particularly in the strong gravitational region near the event horizon, leading to changes in orbital stability and energy requirements. Therefore, studying the orbital dynamics around a spinning \textcolor{black}{charged} QCBH contributes to a deeper understanding of the motion of particles in stable and unstable circular orbits, the accretion process, and the physical properties of compact objects.
\begin{figure}[H]
\centering
\includegraphics[width=85mm,height=60mm]{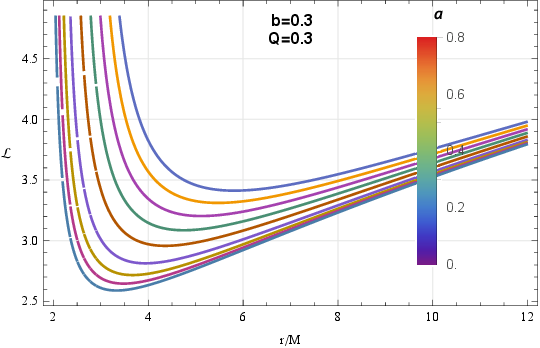}
\includegraphics[width=85mm,height=60mm]{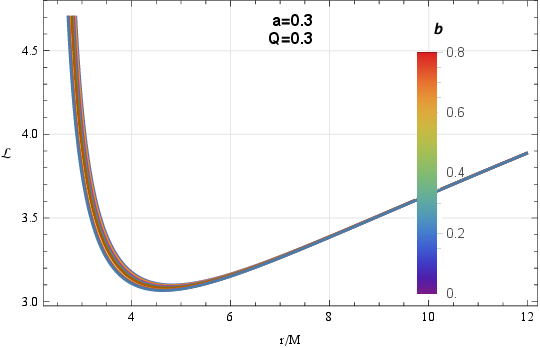}
\includegraphics[width=85mm,height=60mm]{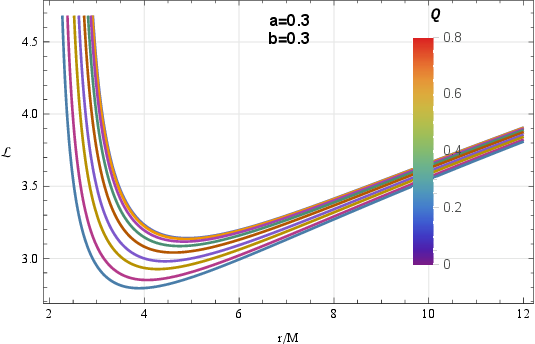}
\caption{Angular momentum $\mathcal{L}$ of test particles orbiting the \\ \phantom{FIG.2.} spinning \textcolor{black}{charged} QCBH.}
\label{fig.2}
\end{figure}
The motion of a neutral particle in a charged QC spacetime can be described in terms of a Hamiltonian \cite{Ashraf2025,Mustafa2025}. Under this framework, the relativistic Hamiltonian can be expressed as:
\begin{equation}
H=\frac{1}{2}g^{\alpha\beta}P_{\alpha}P_{\beta}+\frac{1}{2}m^2.
\label{6}
\end{equation} 
Here $m$ is the mass, while $P^{\gamma}=mu^\gamma$ is the four-momentum with $u^\gamma=dx^\gamma/d\tau$ being the four-velocity, and $\tau$ corresponding to the proper time. The respective Hamilton's equations can be formed as:
\begin{equation}
\frac{dx^\gamma}{d\zeta}=mu^\gamma=\frac{\partial H}{\partial P_\gamma},~~~\frac{dp_\gamma}{d\zeta}=-\frac{\partial H}{\partial x^\gamma}.
\label{7}
\end{equation}
with $\zeta =\tau/m$ is an evolution parameter. The symmetries of spacetime provide two conserved quantities:  
\begin{equation}
\frac{P_t}{m}=g_{tt}u^t+g_{t\phi}u^\phi=-\mathcal{E},~~~\frac{P_\phi}{m}=g_{\phi \phi}u^\phi+g_{t\phi}u^t=\mathcal{L}.
\label{8}
\end{equation}

Here, $(\mathcal{E}=E/m)$ represents the specific energy, expressed in geometric units as a dimensionless parameter, while $(\mathcal{L}=L/m)$ represents the specific angular momentum of the test particle for a spinning \textcolor{black}{charged} QCBH system; the Hamiltonian (\ref{6}) can be written as:  
 \begin{eqnarray}
H &=& \frac{1}{2r^{4}H_{1}}\big( br^{2}(r^{2}+\mathcal{L}^{2}-a\mathcal{L}\mathcal{E}+a^2\varepsilon^2)\nonumber \\ &+&r^4(-a(Q^2-2r)\mathcal{L}\mathcal{E}-r^4\varepsilon^2+a^2H_2\nonumber\\&+&(r^2+\mathcal{L}^2)H_3)+H_1(H_1P_r^2+r^2P_\theta^2)\big).
\label{9}
\end{eqnarray}  
where
\begin{eqnarray} \nonumber
H_1&=& b+r^2\big(a^2+Q^2+(-2+r)r\big),\nonumber \\
H_2&=&r^2+\big(Q^2-r(2+r)\big)\mathcal{E}^2,\nonumber  \\ 
H_3&=&Q^2+(-2+r)r.\nonumber 
\end{eqnarray}

\begin{figure*} 
\centering 
\includegraphics[width=\hsize]{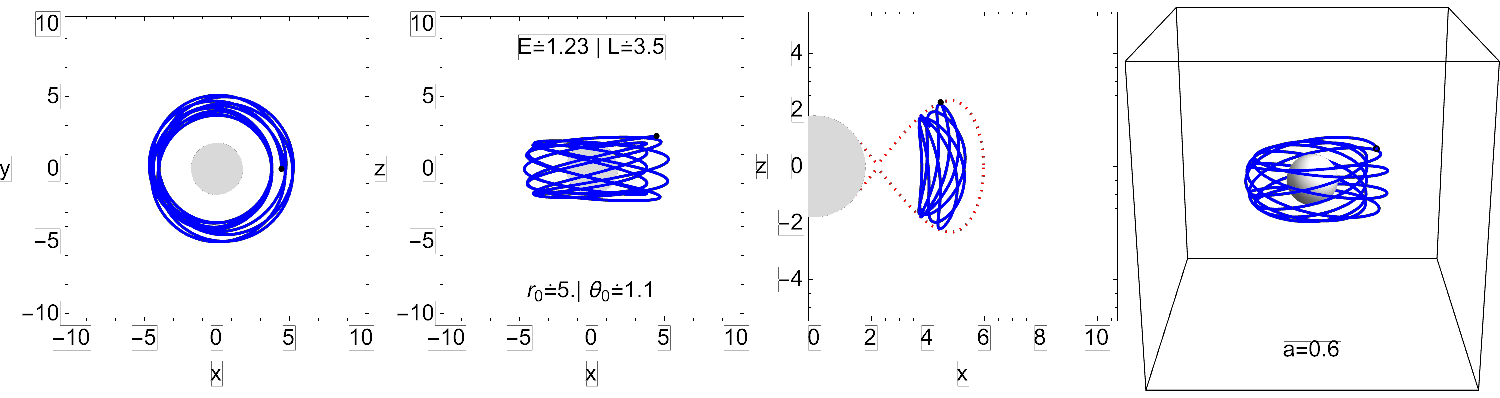}
\includegraphics[width=\hsize]{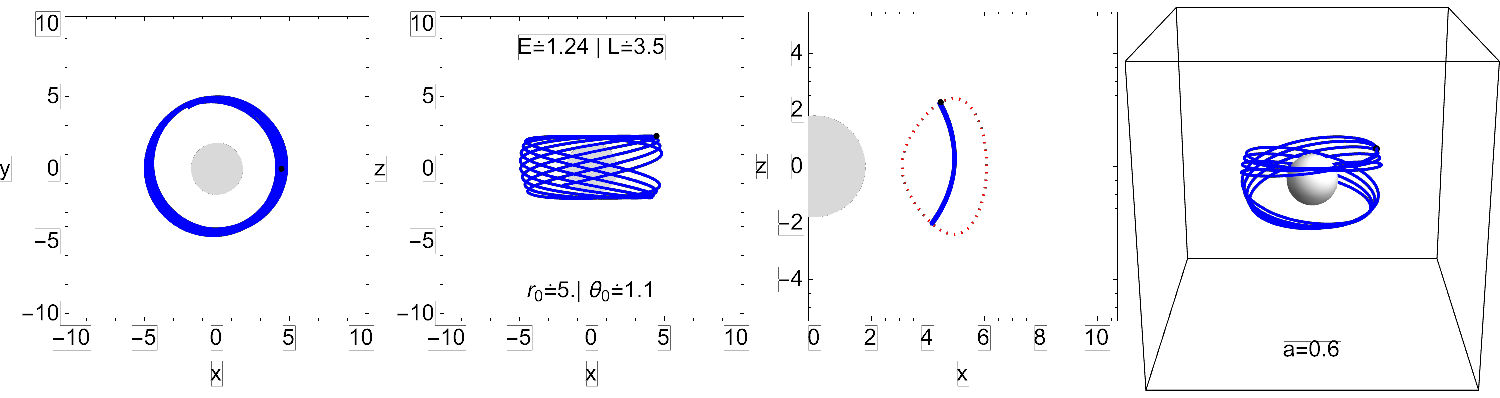}
\includegraphics[width=\hsize]{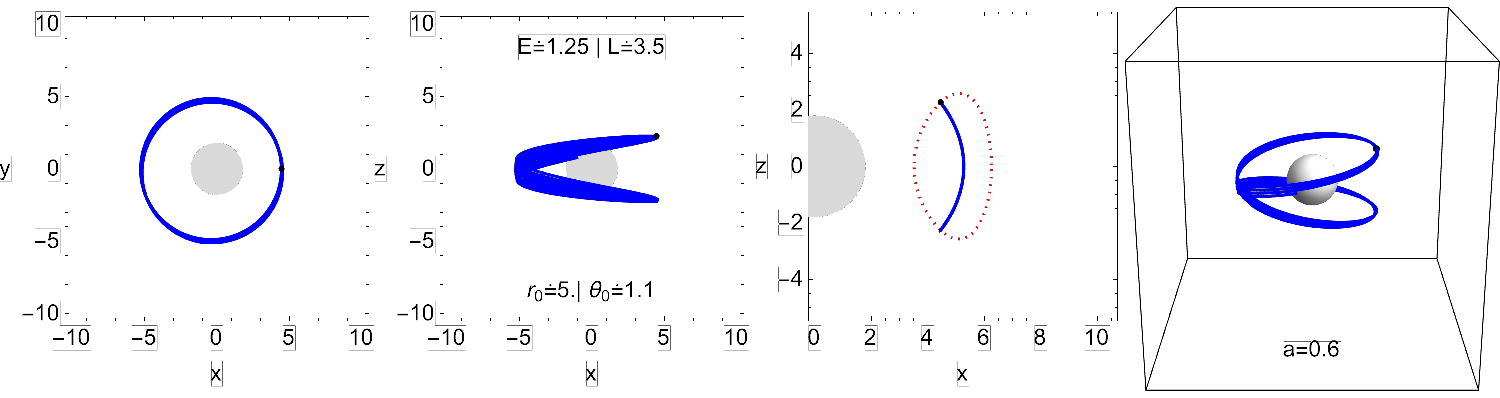}
\caption{\textcolor{black}{Trajectories of test particles around BH by using different values of $Q$: 0.2 (upper panel), 0.4 (middle panel), 0.6 (lower panel) with $M=1,b=0.2$. The trajectories are obtained by numerical integration of the Hamiltonian equations of motion with the initial conditions $r_0=5$, $\theta_0=1.1$, $p_r(0)=p_\theta(0)=0$, and $\phi(0)=0$. The coordinates $(x,y,z)$ are
auxiliary pseudo-Cartesian coordinates defined through 
$x=r\sin\theta\cos\phi$, $y=r\sin\theta\sin\phi$, and
$z=r\cos\theta$, and are used solely for visualization of the motion.}}\label{fig:3}
\end{figure*}

\begin{figure*} 
\centering 
\includegraphics[width=\hsize]{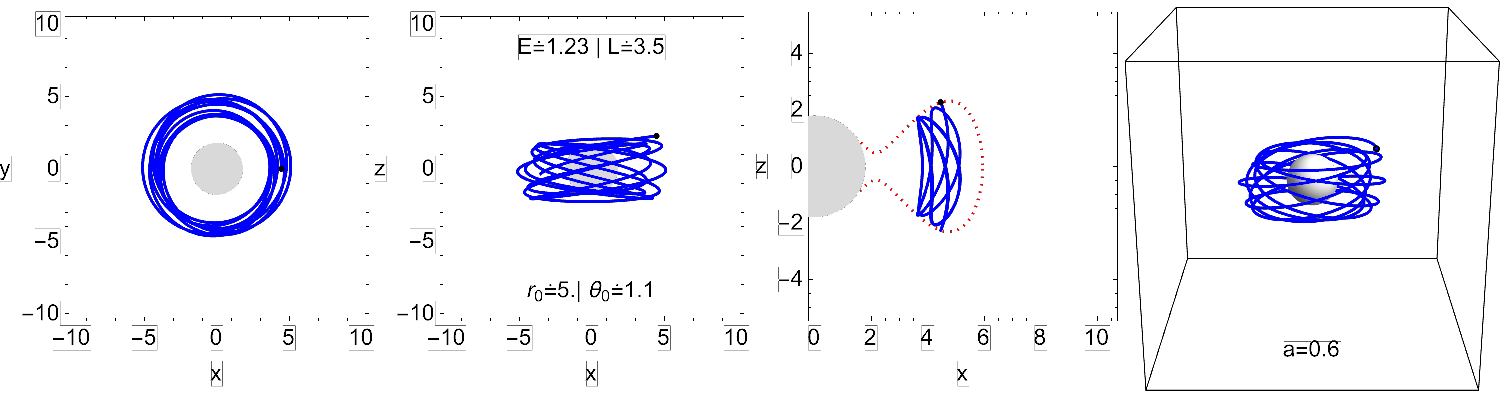}
\includegraphics[width=\hsize]{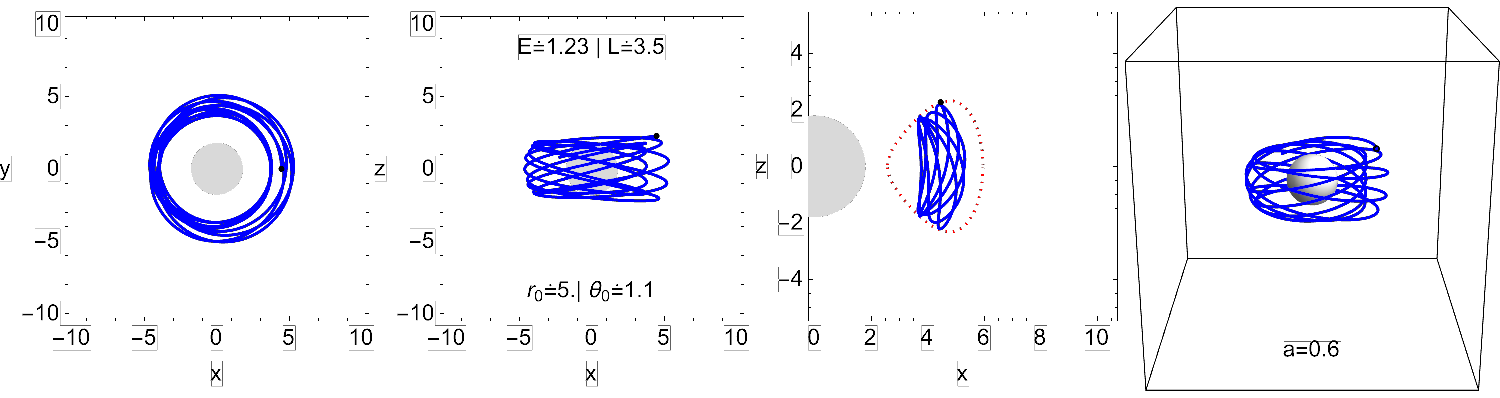}
\includegraphics[width=\hsize]{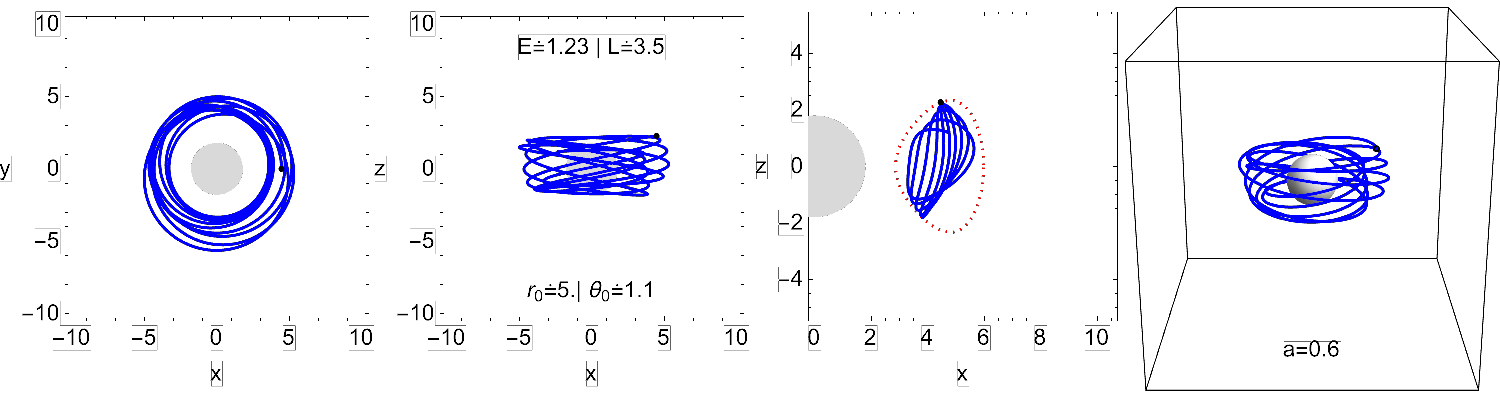}
\caption{\textcolor{black}{Trajectories of test particles around BH using different values of $b$ as 0.4 (upper panel), 0.8 (middle panel), 1.2 (lower panel) with $M=1, Q=0.3$. }
}\label{fig:4}
\end{figure*}
The specific energy and the corresponding specific angular momentum for particles moving in the gravitational field of a spinning \textcolor{black}{charged} QCBH are determined as follows:
\begin{eqnarray} \nonumber
\mathcal{E} &=& \frac{1}{{\mathcal{E}_3}{\mathcal{E}_4}{\sqrt{\mathcal{E}_5}}}\big( r^3(b+r^2(Q^2+(-2+r)r))\\&-& a(b+Q^2-2r)r^2\big)\sqrt{\mathcal{E}_1}+\mathcal{E}_2.  
\label{10}
\end{eqnarray}
\begin{eqnarray}
 \mathcal{L} &=& \frac{1}{{\mathcal{E}_3}{\mathcal{E}_4}{\sqrt{\mathcal{E}_5}}} \big( \sqrt{\mathcal{E}_1}{a^2(r^2(r(r+2)-Q^2)-b)} \nonumber\\ &+& {ar(a^2(2b+r^2(Q^2-r))+3br^2} \nonumber \\ &+& {3br^2+r^4(2Q^2-3r))}\big). 
 \label{11}
\end{eqnarray}
where
\begin{eqnarray}
 \mathcal{E}_1 &=& -2b+r^2(-Q^2+r), \nonumber \\
 \mathcal{E}_2 &=& a^2\big(2br+(Q^2-r)r^3\big), \nonumber\\
 \mathcal{E}_3 &=& r^6+a^2\big(2b+(Q^2-r)r^2\big), \nonumber \\
 \mathcal{E}_4 &=& r^7+{\mathcal{E}}_2, \nonumber \\
 \mathcal{E}_5 &=& r^2\big(2a^3{\mathcal{E}}^\frac{3}{2}_1r+a^2\big(-6b^2+br^2(3r(2r+3)\nonumber\\&-&7Q^2)\big) +r^4(-2Q^4+Q^2r(3r+5)-3r^2(r\nonumber\\&+&1))\big) +2a\sqrt{{\mathcal{E}}_1}r^3\big(r^2(3r-2Q^2)-3b\big)+3br^6\nonumber\\&+&r^8\big(2Q^2+(r-3)\big).\nonumber
\end{eqnarray}

\begin{table}[ht]
\caption{Effect of spin parameter, quantum correction parameter, and \textcolor{black}{electric} charge parameter on the energy evolution $\mathcal{E}$.}
\label{tab1}
\centering
\scriptsize
\renewcommand{\arraystretch}{1.3}

\begin{tabular}{|c|c|c|}
\hline
\textbf{Parameter} &
\textbf{Values} &
\textbf{Interpretation} \\
\hline

$a$ &
\begin{tabular}[c]{@{}c@{}}
0.00, 0.15, 0.25, \\
0.35, 0.50, \\
0.63, 0.70
\end{tabular}
&
\begin{tabular}[c]{@{}c@{}}
Stronger rotational effects\\
lead to deeper energy profiles
\end{tabular}
\\
\hline

$b$ &
\begin{tabular}[c]{@{}c@{}}
0.00, 0.10, 0.20,\\
0.30, 0.40, \\
0.50, 0.60
\end{tabular}
&
\begin{tabular}[c]{@{}c@{}}
Modifies the spacetime geometry\\
and affects the stable orbital region
\end{tabular}
\\
\hline

$Q$ &
\begin{tabular}[c]{@{}c@{}}
0.10, 0.20, 0.30, \\
0.40, 0.50, \\
0.60, 0.66
\end{tabular}
&
\begin{tabular}[c]{@{}c@{}}
\textcolor{black}{BH charge indirectly influencing} \\ \textcolor{black}{ the motion of neutral test particles}
\end{tabular}
\\
\hline
\end{tabular}
\end{table}
\textcolor{black}{Figure \ref{fig.1} shows the specific energy $\mathcal{E}$ of a test particle orbiting a spinning charged QCBH as a function of $a$, $b$, and $Q$. These parameters alters the depth and radial location of the energy, with strong effect near the BH. In the upper panel, $b=0.2$ and $Q=0.2$ remain constant, while $a$ varies. Increasing the spin parameter significantly reduces the minimum value of the specific energy curve and shifts it to a smaller radial distance. This indicates that the spin effect is particularly significant in the strong field region near the BH. At larger radial distances, the energy curves gradually converge, indicating that the effect of spin weakens.
\\ In the middle panel, $a=0.5$ and $Q=0.5$ are fixed, while the quantum correction parameter $b$ varies. Compared to the spin parameter $a$, the change in $b$ has a relatively small impact on the specific energy distribution, with the difference mainly concentrated near the energy minimum. In the lower panel, $a=0.4$ and $b=0.4$ are fixed, while $Q$ increases. The electric charged parameter $Q$ changes both the depth and radial position of the energy minimum; generally, an increase in $Q$ leads to a higher energy minimum in the panel. In both panels, the curves converge with increasing $r/M$, indicating that the effects of the quantum correction and charge mainly operate in the strong field region and become less pronounced at larger radial distances.} 
\\ \textcolor{black}{Figure \ref{fig.2} shows the effects of the parameters $a$, $b$, and $Q$ on the angular momentum $\mathcal{L}$. Variations in these parameters affect the magnitude and radial position of the minimum value of $\mathcal{L}$. Among the three parameters, $b$ exhibits relatively small variation over the range considered. The parameters effects are more pronounced in the strong field region near the BH. As $r/M$ increases, the curve converge, indicating that these effects gradually weaken over larger radial distances. If $b=0.3$ and $Q=0.3$ remain constant, while $a$ changes. Increasing the spin parameter reduces the minimum of the angular momentum distribution curve and shifts it to a smaller radial distance. The separation between the curves is most significant in the strong-field region, indicating that the effect of rotation on $\mathcal{L}$ is stronger near the BH. In the middle panel, $a=0.3$ and $Q=0.3$ are fixed, while $b$ varies. The curves remain close together, indicating that the change in $b$ has only a relatively small effect on the angular momentum distribution. This effect is mainly visible near the minimum value, while at larger radial distances, the curves are almost indistinguishable. Therefore, among the three parameters considered, $b$ has a relatively weak effect on $\mathcal{L}$ for the range of parameters shown. In the figure below, $a=0.3$ and $b=0.3$ remain constant, while $Q$ changes. Increasing the charge parameter alters the minimum values of $\mathcal{L}$ and its radial position. The curve in the figure shows a clear separation in the strong field region, which gradually decreases as $r/M$ increases. Therefore, $a$ and $Q$ have a more significant effect on the relative angular momentum than $b$, especially near the BH, while the effects of all three parameters become less pronounced as the radial distance increases.}
\\ The angular momentum $\mathcal{L}$ steadily increases with the radial distance $r$, and approaches the Keplerian limit $\mathcal{L}\propto \sqrt{Mr}$ at large distance. Furthermore, QC spacetime requires a higher angular momentum than Kerr-Newman black holes at equivalent radii, particularly in the $3M<r<10M$ range \cite{Mustafa2025}. These energy and angular momentum properties play a crucial role in accretion disk dynamics, determining the location of the ISCO, controlling the efficiency of the energy extraction process, and shaping the observable emission spectrum of the accretion disk. The observed perturbation from Kerr-Newman suggests potentially observable effects of the interplay between quantum corrections and electric charge in astrophysical BH systems. Such features may provide a means to distinguish QC spacetimes from charged spinning models through future high-precision observations.
Furthermore, the angular momentum required for a QC spacetime is systematically higher than that of the Kerr-Newman BH of the same radius \cite{Mustafa2025}. These energy and angular momentum characteristics directly affect accretion disk physics, specifically influencing the position of the innermost stable circular orbit (ISCO), the efficiency of the energy extraction mechanism, and the spectral characteristics of accretion disk radiation. The observed deviation from Kerr-Newman predictions \textcolor{black}{indicates characteristic effects of the quantum corrections and charge parameters. However, these differences can be distinguished observationally requires the specific comparison to the astrophysical data}.

\textcolor{black}{The particle trajectories are obtained by numerically integrating the Hamiltonian equations of motion using Mathematica's \texttt{NDSolve}. We employ the \texttt{StiffnessSwitching} algorithm with a ninth-order explicit Runge-Kutta method and Hamiltonian projection, using a starting step size of \(10^{-4}\). The initial conditions are chosen as
$
r(0)=r_0,\qquad \theta(0)=\theta_0,\qquad
p_r(0)=p_\theta(0)=0,\qquad \phi(0)=0.
$
The specific energy \(E\) is obtained from the effective potential, whereas the specific angular momentum \(L\) is fixed. The integration is stopped when the particle reaches the horizon or \(r=1000\). As a representative case, we use \(a=0.6\), \(Q=0.4\), \(b=0.2\), \(r_0=5\), \(\theta_0=1.1\), \(L=3.5\), and \(\tau_{\rm end}=200\).}
Figures~\ref{fig:3} and~\ref{fig:4} illustrate the three-dimensional trajectories of neutral test particles around the spinning \textcolor{black}{charged} QCBH, highlighting the effects of the electric charge parameter $Q$ and the quantum correction parameter $b$, respectively. In Fig.~\ref{fig:3}, the upper, middle, and lower panels correspond to $Q=0.2$, $0.4$, and $0.6$, while keeping $M=1$ and $b=0.2$ fixed. As the electric charge increases, the particle trajectories become progressively more distorted and exhibit stronger orbital precession. The enhanced electromagnetic contribution modifies the effective spacetime geometry, producing larger deviations from closed bound orbits and altering the orbital structure, particularly in the strong-field region near the event horizon. Likewise, Fig.~\ref{fig:4} demonstrates the influence of the quantum correction parameter by considering $b=0.4$, $0.8$, and $1.2$ with fixed $M=1$ and $Q=0.3$. Increasing the quantum correction parameter noticeably changes the orbital morphology, leading to more pronounced relativistic precession and significant deviations from the classical Kerr-Newman trajectories. These quantum-induced modifications become increasingly important in the vicinity of the BH, where the curvature of spacetime is strongest. Overall, both the electric charge and the quantum correction parameter substantially influence the orbital dynamics, indicating that the combined effects of electromagnetic interactions and quantum corrections produce \textcolor{black}{significant theoretical} changes in the trajectories of particles orbiting the spinning \textcolor{black}{charged} QCBH.

\vspace{-0.7cm}
\section*{B.~~~ Effective Potential}
\begin{figure*}[t]
    \centering
    \includegraphics[width=0.32\textwidth]{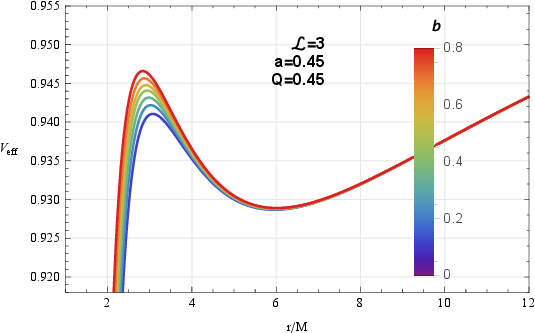}
    \hfill
    \includegraphics[width=0.32\textwidth]{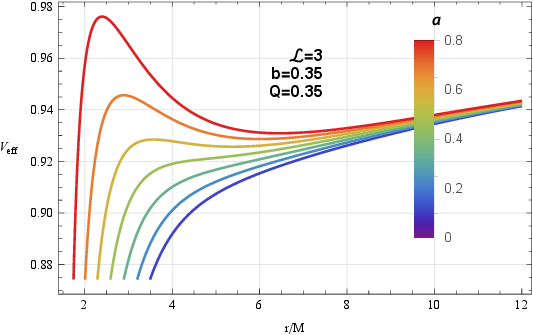}
    \hfill
    \includegraphics[width=0.32\textwidth]{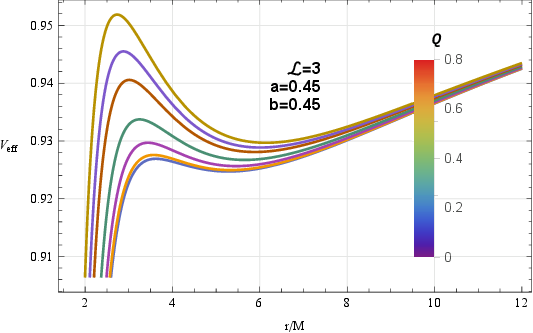}
    \caption{Effective potential for test particles around spinning \textcolor{black}{charged} QCBH.}
    \label{fig.5}
\end{figure*}

\vspace{-0.5cm}
The effective potential framework describes the motion of the particle in curved spacetime and essentially is based on the normalization condition for a massive particle, ${g}_{\nu\sigma}u^{\nu}u^{\sigma}=-1$. Starting from this condition, a general form of the effective potential can be derived, expressed as: \cite{Misner1973, Chandrasekhar1983}    
\begin{equation}
V_{eff}(r,\theta)=g_{rr}\dot r^{2}+g_{\theta\theta} \dot \theta^{2}.  
\label{12}
\end{equation}
Here, $\dot{r}=dr/d\tau$ and $\dot{\theta}=d\theta/d\tau$ are denoted the derivatives of the radial and polar coordinates with respect to the particle proper time. For the Kerr-Newman-(A)DS BH in $f(R)$ gravity, the effective potential $V_{eff}(r,\theta)$ governing the motion of a neutral massive test particle, can be expressed in term of the spacetime metric components as follows: \cite{Bardeen1972, Wilkins1972}.
 \begin{equation}
 V_{eff}(r,\theta)=\frac{\mathcal{E}^2g_{\phi\phi}+2\mathcal{E}\mathcal{L}g_{t\phi}+\mathcal{L}^2g_{tt}}{g_{t\phi}^2-g_{tt}g_{\phi\phi}}-1.  
 \label{13}
 \end{equation}
 Considered for  spacetime geometry, the explicit form is \cite{Quevedo2011, Kato1990}
\begin{eqnarray}
 V_{eff}(r,\theta) &=& \frac{1}{-r^6+a^2\big(b+r^2\big(Q^2-r(2+r)\big)\big)} \nonumber \\ &\times&  \big(\big(a\big(b+(Q^2-2r)r^2\big){\mathcal{L}}\big)\nonumber\\ &-&r\big(H_1a^2\big(-b+r^2\big(-Q^2+r(2+r)\big)\big) \nonumber\\&+&r^4\big(r^2+{\mathcal{L}}^2)\big)^{\frac{1}{2}} \big).  
 \label{14}
\end{eqnarray}
To evaluate the motion of the test particles, the effective potential energy $V_{eff}(r,\theta)$ provides a convenient and reliable framework for describing their dynamics. It allows analysis of particle trajectories without directly integrating the complete equations of motion\cite{Ryan1995, Levin2009}. For particles moving along circular orbits, the effective potential plays an important role in obtaining the necessary conditions for orbital equilibrium. Restricting the motion to the equatorial plane $(\theta=\pi/2)$, simplifies the analysis while preserving the fundamental physical properties of the system. Under this assumption, the radial motion is controlled by the effective potential. A circular orbit is obtained by requiring that the radial derivative be zero. Furthermore, to ensure equilibrium, the radial derivative of the effective potential must be zero. These two requirements determine the possible radii of the circular motion. Therefore, the conditions for a circular orbit can be written as follows:
\begin{equation}
V_{eff}(r)=0,~~~\frac{\partial V_{eff}(r)}{\partial r}=0.  
\label{15}
\end{equation}
These requirements are important for defining the ISCO and other fundamental radii \cite{Abramowicz2013}. The effective potential $V_{eff}(r)$ indicates the existence of circular orbits, with minimum value, corresponding to stable configurations and maximum values to unstable configurations. Unlike Newtonian gravity, the orbital positions of these relativistic systems are strongly dependent on angular momentum $\mathcal{L}$ and specific BH parameters. 
\cite{Bardeen1972}. 
For spinning \textcolor{black}{charged} QCBHs, the stability of circular orbits is influenced by $a$, $b$, and $Q$. These parameters collectively alter the spacetime geometry and affect the motion of the \textcolor{black}{neutral} test particle. The position of the innermost stable circular orbit (ISCO) can be obtained by simultaneously satisfying the conditions for circular motion and critical stability. Solving these equations determines the transition between stable and unstable orbits. Therefore, ISCO provides a fundamental method for measuring orbital stability and plays an important role in understanding the particle's motion near spinning \textcolor{black}{charged} QCBHs.
\begin{equation}
\frac{\partial{V}_{eff}}{\partial r}=0,~~~\textcolor{black}{\frac{\partial^2{V_{eff}}}{\partial r^2}\leq0}.  
\label{16}
\end{equation}
\textcolor{black}{Table \ref{tab:isco} lists the ISCO radii for representative parameter values. The table also includes  Kerr and Kerr-Newman limits to demonstrate that the corresponding classical cases can be recovered and  to quantify the correction introduced by the quantum correction parameter $b$.}
\begin{table}[htbp]
\centering
\caption{ISCO radii for representative parameter values with $M=1$.}
\label{tab:isco}
\renewcommand{\arraystretch}{1.9}
\setlength{\tabcolsep}{5pt}
\footnotesize

\begin{tabular}{|c|c|c|}
\hline
\textbf{BH Model} & \textbf{Values} & 
\textbf{$r_{\rm ISCO}/M$} \\
\hline

Kerr
& $a=0.55,\ b=0,\ Q=0$
& 4.03432 \\
\hline

uncharged QC
& $a=0.55,\ b=0.45,\ Q=0$
& 3.84568 \\
\hline

Kerr--Newman
& $a=0.55,\ b=0,\ Q=0.45$
& 3.61967 \\
\hline

charged QC 
& $a=0.55,\ b=0.20,\ Q=0.45$
& 3.50705 \\
\hline

charged QC
& $a=0.55, \ b=0.45, \ Q=0.45$
& 3.33661 \\
\hline

\end{tabular}
\end{table}
\textcolor{black}{The results show that, in the Kerr limit, the radius of the ISCO is $r_{ISCO}=4.03432M$, while in the corresponding Kerr-Newman configuration with $a=0.55$ and $Q=0.45$, the radius of the ISCO is $r_{ISCO}=3.61967M$. Also, with $a=0.55$ and $Q=0.45$ fixed, increasing the quantum correction parameter $b$ from $0$ to $0.20$ and $0.45$ reduces the radius of the ISCO from $3.61967M$ to $3.50705M$ and $3.33661M$, respectively. Therefore, within the considered parameter range, the quantum correction shifts the prograde ISCO inward relative to the Kerr-Newman case.}
\textcolor{black}{\\Figure \ref{fig.5} illustrates the effective potential $V_{eff}$ of a neutral test particle orbiting a spinning charged QCBH with fixed angular momentum $\mathcal{L}=3$. In the left panel, with a fixed $a=0.45$ and $Q=0.45$, changes in $b$ have only a small effect on the effective potential distribution, mainly affecting the area near the potential peak in the strong field region. In the middle panel, with a fixed $b=0.35$ and $Q=0.35$, an increase in a significantly alters the height and radial structures of the potential peak, indicating that spin has a significant impact on the motion of particles near the BH. In the right panel, with a fixed $a=0.45$ and $b=0.45$, changes in $Q$ also significantly alter the height and shape of the effective potential peak. In all three panels, the distance between the curves decreases with increasing radial distance, indicating that the effects of $a$, $b$, and $Q$ are mainly concentrated in the strong field region and gradually weaken with distance from the BH}.
\section*{C.~~~ Effective Force }
The effective force on the particle motion is a fundamental physical quantity for understanding the spacetime of a spinning \textcolor{black}{charged} QCBH. It determines whether the gravitational interaction acting on the particle is primarily attractive (pulling the particle toward the QCBH) or repulsive (pushing the particle far from the central body). Therefore, the effective force provides a valuable perspective for understanding the dynamical behavior of particles moving near a spinning \textcolor{black}{charged} QCBH. Studying this force helps identify regions where competing interactions dominate and elucidates the conditions under which stable or unstable motion may occur. In this study, we investigate the effective force to better understand the comprehensive effects of BH parameters on particle dynamics. The radial effective force that governing the motion of the test particles can be expressed as \cite{Mustafa2025, Ashraf2025}:
\begin{equation}
F=-\frac{1}{2}\frac{dV_{eff}}{dr}.  
\label{17}
\end{equation}
\begin{figure}
\centering
\includegraphics[width=85mm,height=60mm]{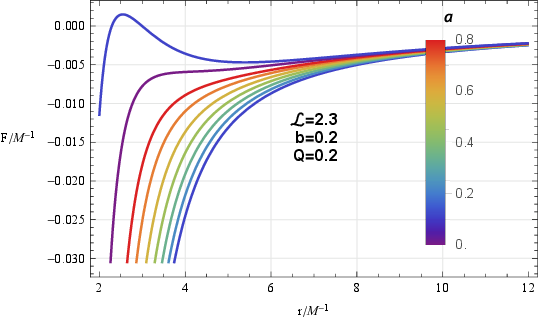}
\includegraphics[width=85mm,height=60mm]{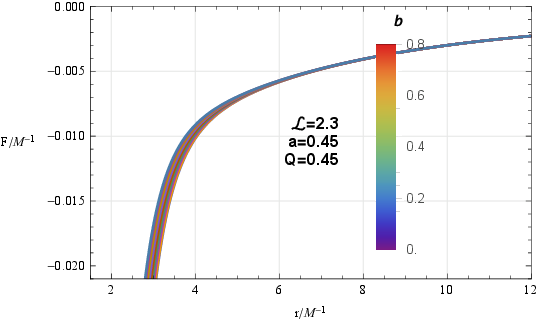}
\includegraphics[width=85mm,height=60mm]{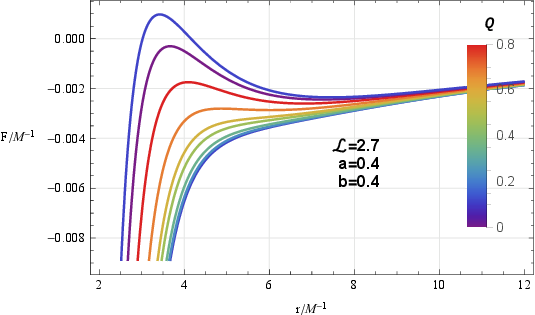}
\caption{Effective force for test particles moving around \\ \phantom{FIG.6.}~ spinning \textcolor{black}{charged} QCBH.}
\label{fig.6}
\vspace{0.5cm}
\end{figure}
\begin{table}[ht]
\caption{Effect of $a$, $b$, and $Q$ parameters on the angular momentum evolution $\mathcal{L}$.}
\label{tab2}
\centering
\scriptsize
\renewcommand{\arraystretch}{1.3}

\begin{tabular}{|c|c|c|}
\hline
\textbf{Parameter} &
\textbf{Values} &
\textbf{Interpretation} \\
\hline

$a$ &
\begin{tabular}[c]{@{}c@{}}
0.00, 0.10, 0.20, \\
0.30, 0.40, 0.50, \\
0.56, 0.60, 0.63
\end{tabular}
&
\begin{tabular}[c]{@{}c@{}}
frame-dragging effects\\
angular momentum for stability
\end{tabular}
\\
\hline

$b$ &
\begin{tabular}[c]{@{}c@{}}
0.00, 0.10, 0.20,\\
0.30, 0.40, 0.50,\\
0.60, 0.65, 0.69
\end{tabular}
&
\begin{tabular}[c]{@{}c@{}}
Modifies the spacetime geometry\\
and affects orbital angular \\
momentum
\end{tabular}
\\
\hline

$Q$ &
\begin{tabular}[c]{@{}c@{}}
0.00, 0.10, 0.20, \\
0.30, 0.40, 0.50 \\
0.57, 0.65, 0.70
\end{tabular}
&
\begin{tabular}[c]{@{}c@{}}
\textcolor{black}{BH charge modifies the geometry and}\\ \textcolor{black}{affects orbital angular momentum}
\end{tabular}
\\
\hline

\end{tabular}
\end{table}
\textcolor{black}{In Figure \ref{fig.6} the upper panel, with $\mathcal{L}=2.3$, $b=0.2$ and $Q=0.2$ fixed, increasing the value of $a$ significantly alters the effective force distribution in the strong field region. When the value of $a$ is sufficiently large, the effective force becomes positive within a finite radial range, indicating a local outward contribution of the effective force, while the effective force distribution on other regions is mainly negative. In the middle  panel, with $\mathcal{L}=2.3$, $a=0.45$ and $Q=0.45$ fixed, the change in effective force caused by variations in the value of $b$ is relatively small, and the effective force remains negative within the radial range. In the lower panel, with $\mathcal{L}=2.7$, $a=0.4$, and $b=0.4$ fixed, increasing the value of $Q$ significantly alters the magnitude of the effective force and the radial structure, and when the value of $Q$ is sufficiently large, the effective force can become positive within a finite radial range. In all three panels, the separation between the curves decrease with increasing radial distance, indicating that the effects of $a$, $b$, and $Q$ are mainly significant in the strong field region and gradually weaken with the distance from the BH}. 
\section* {III.~~~EXAMINING THE HARMONIC OSCILLATIONS AS PERTURBATION OF CIRCULAR ORBITAL MOTION}
We investigate the deviations from a stable circular orbit in the equatorial plane to explore the oscillation behavior of neutral particles. When the test particle deviates slightly from its equilibrium state, it undergoes a cyclic motion, which can be described as a combination of radial and vertical harmonic oscillations. The fundamental frequencies of these oscillations are observed in a co-moving frame and can be obtained by the following formulas \cite{Ashraf2025, Mustafa2025} 
\begin{equation}
\omega_r^2=\frac{-1}{2g_{rr}} \frac{\partial^2{V_{eff}(r,\theta)}}{\partial r^2}.  
\label{18}
\end{equation}
\begin{equation}
\omega_\theta^2=\frac{-1}{2g_{\theta\theta}} \frac{\partial^2{V_{eff}(r,\theta)}}{\partial \theta^2}.    
\label{19}
\end{equation}
\begin{equation}
\omega_\phi=\frac{d\phi}{dr}. 
\label{20}
\end{equation}
Examining the fundamental frequencies $(\omega_r, ~\omega_\theta, ~\omega_\phi)$ and their ratios provides insight into the structure of rotational motion around a stable circular orbit. In the Newtonian limit, these three frequencies overlap  $(\omega_r = \omega_\theta = \omega_\phi)$, resulting in a simple elliptical trajectory around a spherical mass. In contrast, for a Schwarzschild BH, this degeneracy is eliminated, resulting in a clear frequency separation that produces different relativistic effects: First, the periapsis of the apex is determined by $(\omega_\phi - \omega_r)$, Second, while the precession of the orbital plane is determined by $(\omega_\phi - \omega_\theta)$. Both effects become increasingly significant as the particle approaches the BH.      
\section*{A~~~MEASUREMENT OF FREQUENCIES WITH DISTANT OBSERVER}
The angular frequency denoted by $\omega_{\alpha}$, is represented by equations (\ref{18}), (\ref{19}), (\ref{20}) can be mapped to the frequencies measured by a distant static observer, the corresponding frequencies are denoted by $\Omega_{\alpha}$ can be written as:\cite{Ashraf2025, Mustafa2025}     
\begin{equation}
\textcolor{black}{\Omega_\alpha=\omega_\alpha \bigg(\frac{d\tau}{dt}\bigg)}. 
\label{21}
\end{equation}
Since the signal undergoes a gravitational redshift as it propagates through curved spacetime, the observed frequencies differ from the local frequencies. Therefore, the frequencies measured at infinity are obtained by multiplying the local frequencies by the redshift factor $dt/d\tau$. This relationship can be expressed as: \cite{Misner1973, Chandrasekhar1983}
\begin{equation}
\frac{dt}{d\tau}=-\frac{Eg_{\phi \phi}+Lg_{t\phi}}{g_{tt}g_{\phi \phi}-g_{t\phi}^2}.  
\label{22}
\end{equation}
The frequencies associated with small harmonic oscillations can be expressed in physical units, measured by an observer far from the EMd QCBH. To obtain the corresponding dimensionless quantity, the frequency is normalized by multiplying by a factor  $c^3/GM$, where $c$ is speed of light, $G$ is the gravitational constant, and $M$ is the QCBH mass. This normalization method provides a convenient framework for comparing the oscillation frequencies of QCBH with different masses. It also establishes a direct link between theoretical calculation and observable astrophysical quantities. Therefore, the oscillation frequencies of the particles detected by distant observers can be determined based on the corresponding dimensionless frequencies. \textcolor{black}{In the future}, these physical frequencies can be expressed in hertz $Hz$ for comparison with observational data. The resulting relationship is as follows: 
\begin{equation}
v_j=\frac{1}{2\pi}\bigg(\frac{c^3}{GM}\bigg) \Omega_j[H_z]. 
\label{23}
\end{equation} 
Here $j\in r,\theta,\phi$, where $\Omega_r$ represents the radial frequency, $\Omega_\theta$ represents the vertical frequency, and $\Omega_\phi$ represents the angular frequency measured by an observer at a distance. For a spinning \textcolor{black}{charged} QCBH, the corresponding expression for $\Omega_\alpha$ is as follows:
\begin{eqnarray}
    \Omega_r^2&=&\frac{1}{\Omega_7}(2a ~\Omega_2~ r^2+\Omega_1+\Omega_3+\Omega_4+   b r^4 ~\Omega_5~\mathcal{L} \mathcal{E}\nonumber \\ &+&\Omega_6~\mathcal{E}^2)
   \label{24} 
\end{eqnarray}
\begin{eqnarray}
\Omega_\theta^2 &=& \frac{1}{r^2a\big(b+(Q^2-2r)r^2\big){\mathcal{L}+{\Omega_9}}} \nonumber \\ &\times& \big(H_1r^2(b+r^2H_3){\mathcal{L}+{\Omega_8}-r^3(2+r){\mathcal{E}^2}}\nonumber\\ &+& Q^2(2{\mathcal{L}^2+r^2{\mathcal{E}^2})}\big).
\label{25}
\end{eqnarray} 
\begin{eqnarray}
\Omega_\phi&=& \frac{1}{r^6+a^2\big(2b+(Q^2-r)r^2\big)}\times2ab+a(Q^2\nonumber\\&-&r)r^2+r^3 \big({-2b-Q^2r^2+r^3}\big)^{\frac{1}{2}}.
\label{26}
\end{eqnarray} 
Here, $\Omega_1$, $\Omega_2$, $\Omega_3$, $\Omega_4$, $\Omega_5$, $\Omega_6$ $\Omega_7$ are given in Appendix~\ref{Radial} and $\Omega_8$, $\Omega_9$ are given in Appendix ~\ref{vertical}.
Figure \ref{fig.7} shows the radial oscillation frequency $v_j$ measured by a distant observer, corresponding to the tiny harmonic oscillation of neutral particles orbiting a spinning \textcolor{black}{charged QCBH. For the case of non-spinning $(a=0)$, the vertical and angular frequencies are the same.} As the spin parameter $a$ increases, the frequency distribution gradually shifts outward, away from the event horizon, reflecting the stronger influence of spinning on orbital dynamics. This outward shift indicates that a spinning \textcolor{black}{charged} QCBH alters the location of characteristic oscillation frequencies significantly affect on the particle motion in the surrounding spacetime. 
\begin{figure*}[t]
\centering
\includegraphics[width=0.32\textwidth]{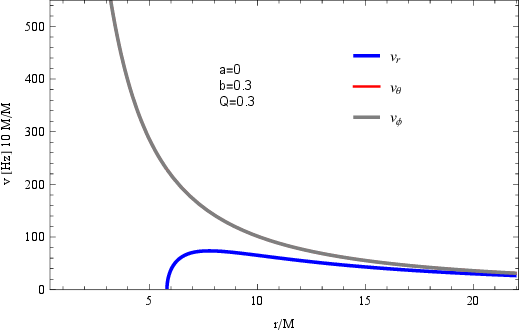} \hfill
\includegraphics[width=0.32\textwidth]{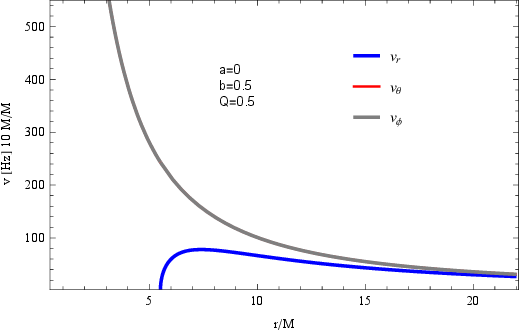} \hfill
\includegraphics[width=0.32\textwidth]{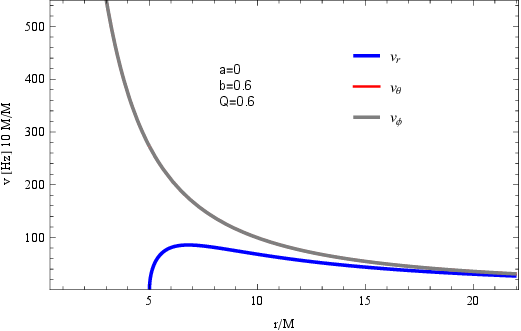} \\[3mm]
\includegraphics[width=0.32\textwidth]{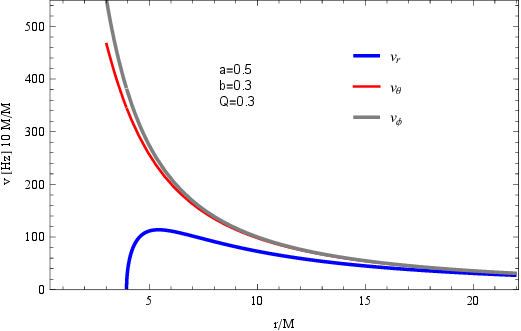} \hfill
\includegraphics[width=0.32\textwidth]{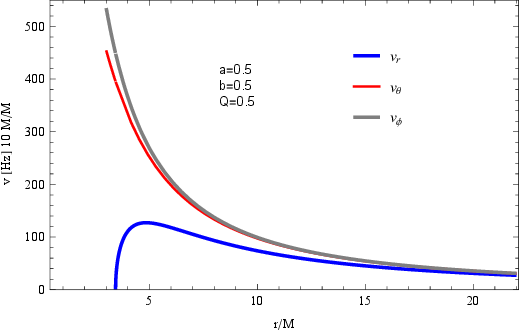} \hfill
\includegraphics[width=0.32\textwidth]{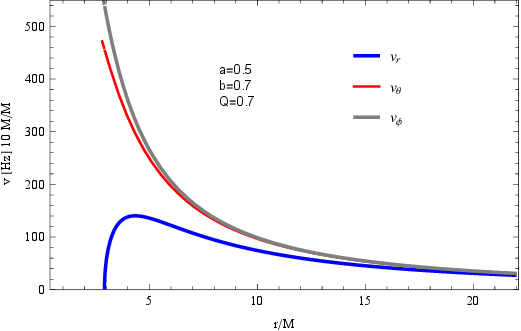} \\[3mm]
\includegraphics[width=0.32\textwidth]{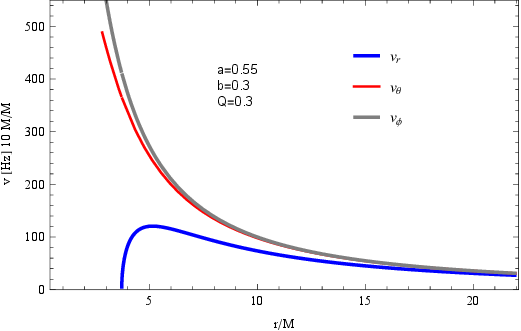} \hfill
\includegraphics[width=0.32\textwidth]{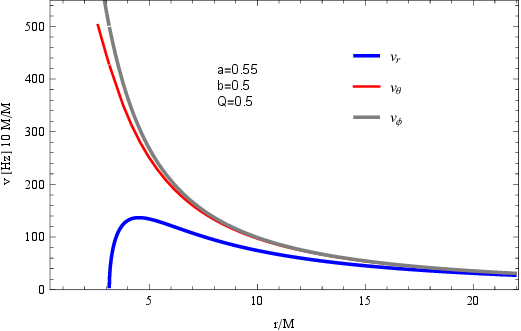} \hfill
\includegraphics[width=0.32\textwidth]{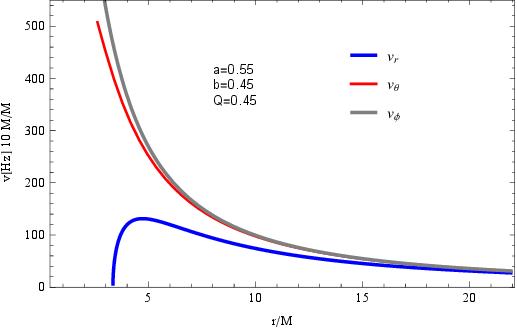} \\[3mm]
\caption{Fundamental frequencies of test particles moving around spinning \textcolor{black}{charged} QCBH.}
\label{fig.7}
\end{figure*}
\section*{B~~~PERIAPSIS and LENSE-THIRRING PRECESSION IN SPACETIME}
In this section \cite{Ashraf2025,Mustafa2025}, we investigate the Peristellar precession frequencies and Lense-Thirring precession frequencies of a neutral test particle orbiting a spinning \textcolor{black}{charged} QCBH. The analysis focused on a particle moving along a nearly stable circular orbit with small perturbations near its equilibrium orbit. Restricting the motion to the equatorial plane  $(\theta \approx {\pi}/2)$, we examine how BH's spinning affect the orbit dynamics. Small radial or vertical displacements deviating from the circular orbit result in oscillatory motion, characterized by radial $\Omega_r$ and vertical $\Omega_{\theta}$ spinning frequencies. These characteristic frequencies provide crucial information about the stability of the particle's motion near the spinning \textcolor{black}{charged} QCBH. The peristellar precession frequency $(\Omega_p)$ quantifies the gradual spinning of the orbital ellipse within the orbital plane, while the Lense-Thirring precession frequency $\Omega_{LT}$  describes the nodal precession caused by the dragging effects of the spinning spacetime frame. These precession frequencies are directly derived from the fundamental orbital frequency and spinning frequency. Specifically, the precession frequency of the periapsis is determined by the difference between the angular orbital frequency $\Omega_{\phi}$ and the radial spinning frequency $\Omega_r$, while the Lense-Thirring precession frequency is determined by the difference between the angular frequency $\Omega_{\phi}$  and the vertical frequency $\Omega_{\theta}$. 
\textcolor{black}{These physical quantities describe the influence of model parameters on particle dynamics and may have potential observational significance. To determine their ability to distinguish spinning \textcolor{black}{charged} QCBHs from other compact objects, they need to be compared with observational data}.\cite{Ashraf2025,Mustafa2025} 

A test particle orbiting a spinning \textcolor{black}{charged} QCBH has an angular frequency, denoted by $\Omega_\phi$. If the radial frequency and angular frequencies satisfy $\Omega_r=\Omega_{\phi}$, the orbit is completely closed. However, in realistic situations, $\Omega_r<\Omega_\phi$, causing the particle to complete angular rotation during a single radial oscillation. 
Therefore, the closest point (periapsis) gradually advances with each orbital motion. This relativistic orbital shift produces the Periapsis precession frequency, expressed as:         
\begin{equation}
\Omega _p=\Omega_\phi-\Omega_r.  ~~~~~~~~~~~~~~~~~~~~~~~~~~~~~~~~~~~~~~~~~~~  
\label{35}
\end{equation}
Similarly, the vertical rotational frequency satisfies $\Omega_\theta<\Omega_\phi$. Due to the frame-dragging effect caused by the rotation of the spinning \textcolor{black}{charged} QCBH, the orbital plane undergoes slow precession about the spin axis. This relativistic nodal motion causes the orbital plane to rotate slowly over time. Therefore, the Lense-Thirring (nodal) precession frequency is defined as:   
\begin{equation}
\Omega_{LT}=\Omega_\phi-\Omega_\theta.    ~~~~~~~~~~~~~~~~~~~~~~~~~~~~~~~~~~~~~
\label{36}
\end{equation}
The precession frequencies obtained in the spacetime of \textcolor{black}{charged} QCBH exhibit behavior not found in Newtonian gravity. In Newton gravity, the radial and angular frequencies are the same $\Omega_r=\Omega_{\phi}$, resulting in closed elliptical orbits. In contrast, the curvature of relativistic spacetime causes these frequencies to differ, thus producing orbital precession. The periapsis precession frequency $\Omega_p$ characterizes the gradual spinning of the orbital ellipse within the equatorial plane and is an important indicator of relativistic effects.  

\begin{figure}[H]
\centering
\includegraphics[width=85mm,height=60mm]{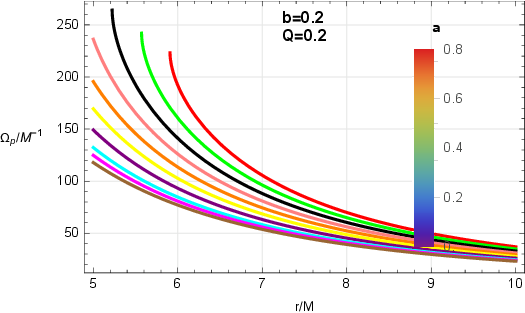}
\includegraphics[width=85mm,height=60mm]{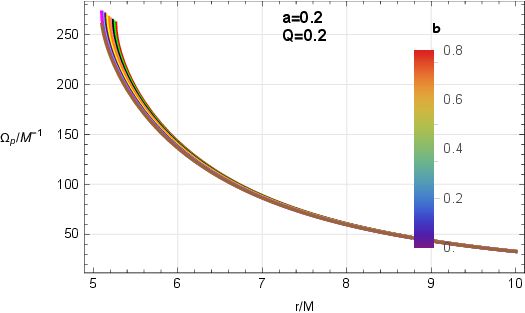}
\includegraphics[width=85mm,height=60mm]{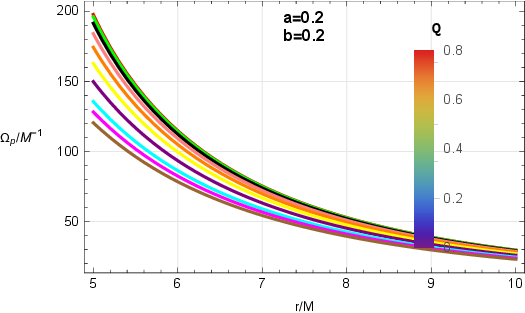}
\caption{Periastron frequency of test particles around spinning \\ \phantom{FIG.8.}~~ \textcolor{black}{charged} QCBH.}
\label{fig.8}
\vspace{0.5cm}
\end{figure}

Figure \ref{fig.8} illustrates the near-star precession frequencies of spinning \textcolor{black}{charged} QCBH as a function of radial coordinate $r$ considering different BH parameter values. The result shows that increasing the QCBH parameters alters the spacetime geometry, leading to a decrease in the near star precession frequency within stable orbital region. Conversely, a larger spinning parameter $a$ enhances the frame-dragging effect and increases the precession frequency, especially near the event horizon.  Furthermore, the precession frequency gradually decreases, increasing the radial distance, becoming negligible in the asymptotic region where the relativistic effects weaken. Comparison with Kerr BH reveals that spinning charged QCBH exhibit unique precession characteristics due to the combined effects of the dilaton field and the QCBH itself. These differences provide valuable insight into the orbital dynamics of test particles and \textcolor{black}{The study provides typical theoretical differences between the rspinning charged QCBH and the standard Kerr spacetime, but their observational distinguishability remains to be determined}.

\begin{figure}[H]
\centering
\includegraphics[width=85mm,height=60mm]{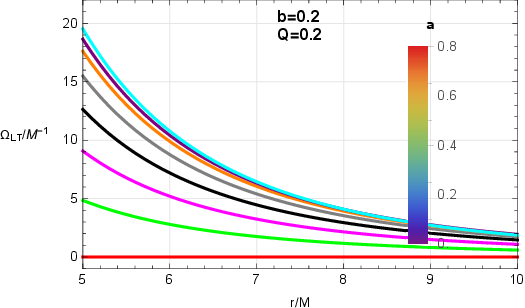}
\includegraphics[width=85mm,height=60mm]{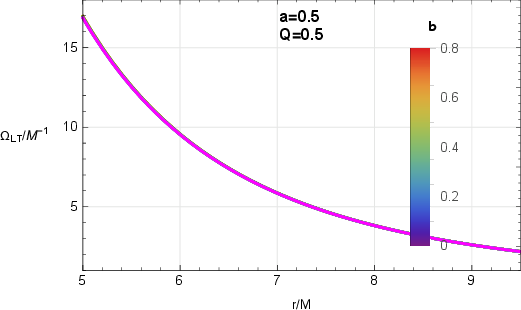}
\includegraphics[width=85mm,height=60mm]{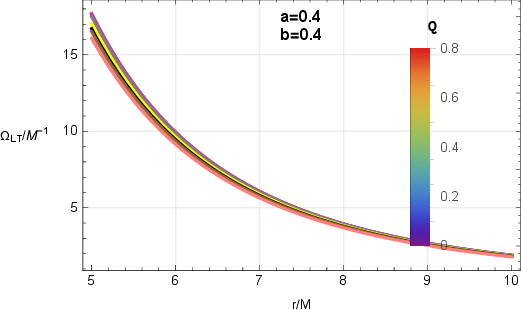}
\caption{Lense-Thirring precession frequency around spinning \\ \phantom{FIG.9.}~~  \textcolor{black}{charged} QCBH.}
\label{fig.9}
\end{figure}

If vertical and angular frequencies are the same $\Omega_\phi=\Omega_\theta$, then Lense-Thirring (nodal) precession vanishes. The Lense-Thirring precession frequency exists only in spinning \textcolor{black}{charged} BH and originates from the frame-dragging effect created by spinning spacetime. It describes the precession of the orbital plane about the spin axis and provides valuable insight into the rotational characteristics of spinning \textcolor{black}{charged} QCBH.
Together, these two precession frequencies provide a useful framework for studying relativistic orbital dynamics and \textcolor{black}{identify theoretically significant features that can be used for observational studies in future research}.
Figure \ref{fig.9} shows the changes in the Lense-Thirring precession frequency around the spinning \textcolor{black}{charged} QCBH. The analysis focuses on the combined effects of $a$, $b$, and $Q$ of the precession motion of nearby test particles. The first panel shows the change in precession frequency with increasing $a$, while the others examine the effects of $b$, and $Q$. The results show that a larger spin enhances the frame-dragging effects, resulting in a higher Lense-Thirring precession frequency near the QCBH. Conversely, increasing the charge or quantum correction parameter weakens the effective spinning effect of spacetime, thereby suppressing the precession frequency. This suppression is more pronounced near the horizon, where strong gravitational effects dominate. Furthermore, the simultaneous presence of charge and quantum correction alters the spacetime geometry, leading to a significant deviation from Kerr behavior. These findings suggest that the spinning parameter primarily enhances the frame-dragging effects, while charge and quantum correction play a role in moderating its amplitude, thus providing the unique character of the spacetime of the spinning charged QCBH. 
\section* {IV.~~~COMPARISON OF RESULTS}
\textcolor{black}{Although the orbital and rotational motions considered in this study are similar to those examined in previous studies \cite{Mustafa2024, Zulqarnain2026, Ashraf2025}, the spacetime and control parameters differ. The variations in orbital motion, rotational frequency, and precession frequency arise from the combined effects of spin $a$, electric charge $Q$, and quantum correction $b$. Specifically, spin has the most significant effect on the precession frequency, while the corrections introduced by $b$ and $Q$ are relatively small within the range of parameters considered}.
The study of physical phenomena near spinning \textcolor{black}{charged} QCBH is crucial for understanding strong gravitational fields and their astrophysical implications. Detailed analysis of the parameters controlling particle motions  provides valuable insight into the dynamics of matter near QCBH. In particular, the theoretical determination of the spinning oscillation frequencies provides an effective framework for exploring orbital dynamics and accretion processes. The combined effects of spinning parameters, electric charged, and quantum correction parameters significantly alter the radial, vertical, and angular oscillation frequencies. These parameters also influence the position and stability of circular orbits, including transitions between stable and unstable orbital configurations. Examining the individual and combined effects of these parameters helps us better understand the spacetime geometry around spinning \textcolor{black}{charged} QCBH. Such analysis are essential for studying relativistic precession and explaining epicyclic oscillations observed in X-ray binaries. Furthermore, they provide important tests for strong-field gravitational theories and establish a direct link between theoretical predictions and astronomical observations. Therefore, the study of spinning oscillations around \textcolor{black}{charged} QCBH plays a vital role in modern relativistic astrophysics and BH physics.    
\section*{CONCLUSIONS}
The EMd framework extends the traditional Einstein-Maxwell theory by introducing a dilaton scalar field that trivially couples with the electromagnetic field when $\gamma=0$. The Spinning charged QCBH solution provides a consistent framework for studying the effects of the dilaton parameter $\gamma$ on particle dynamics and orbital oscillations. Therefore, this framework offers valuable insights into the interactions between gravity, electromagnetism, and scalar fields, making it an important model for studying strong gravitational phenomena and testing related predictions of string theory. 
The outcomes of our discussion can be explained by the way that the spinning, quantum correction and electric charged parameters with neutral test particles alter the energy evolution and angular momentum in unvarying orbits. As shown in Fig.~\ref{fig.1} and Fig.~\ref{fig.2}, higher values of $a$ indicate stronger rotational effects and deeper energy minima for test particle motion and higher spin strengthens frame-dragging effects and rotational influence, allowing particles to orbit closer to the BH, while requiring larger angular momentum for stability. Increasing $b$ slightly modifies the spacetime geometry and shifts the stable orbital region inward the spacetime, but produces only weak effects on orbital angular momentum and stable orbital locations. Higher \textcolor{black}{ values of $Q$} \textcolor{black}{modify the spacetime geometry and consequently affect the motion of neutral test particles} and reduce the minimum energy of the orbit and the effective gravitational attraction of the electric charge weakens and increase the angular momentum required for stable particle motion.
Furthermore, the trajectory analysis presented in Figs.~\ref{fig:3} and~\ref{fig:4} demonstrates that both parameters $Q$ and $b$ significantly modify the geodesic motion of neutral test particles around the spinning charged QCBH. Increasing either parameter enhances the relativistic deformation of particle trajectories and leads to stronger orbital precession, with the most pronounced effects occurring in the strong-gravity region near the event horizon. \textcolor{black}{These deviations provide a characteristic theoretical marker for the spinning \textcolor{black}{charged} QCBH relative to their classical counterparts. We determine that whether these differences can be detected through observation requires quantitative comparison with astrophysical observations}.
Fig.\ref{fig.5} shows the effective potential $\mathcal{E}$ distribution, indicating that the orbital motion of particles around spinning charged QCBH is primarily influenced by $a$, $b$, and $Q$. With increasing spin, the effective potential $\mathcal{E}$ shifts upward due to enhanced spinning and frame-dragging effects, lowering the potential barrier and allowing stable circular orbits to occur closer to the event horizon. The quantum correction parameter $b$ further alters the shape of the potential, with its effects particularly significant in the strong-field region. Furthermore, increasing the $Q$ changes the depth and position of the potential wall through its modification of the spacetime geometry, thus altering the stability of the particle orbit.
Fig.\ref{fig.6} shows the corresponding effective force $\mathcal{L}$ distribution, indicating that these parameters collectively influenced the gravitational force experienced by the particles. The parameter $a$ enhances the spinning contribution of the effective force, while the parameter $b$ introduces additional bias near the event horizon. Moreover, parameter $Q$ significantly alters the magnitude and distribution of the effective force $\mathcal{L}$, leading to significant changes in particle dynamics and orbital stability. These results demonstrate that the intersection between $a$, $b$, and $Q$ plays a fundamental role in determining the effective potential $\mathcal{E}$ and effective force $\mathcal{L}$ in the spacetime of a spinning charged QCBH.  
Our computation of fundamental frequencies of charged QCBH reveals that the combined effects of $a$, $b$, and $Q$ significantly alter the $(v_r)$, $(v_{\theta})$ and $(v_{\phi})$ Fig.~\ref{fig.7}  frequency distributions. These parameters change the spacetime geometry and effective gravitational potential, causing the characteristic frequency distributions to shift towards the event horizon as the quantum correction increases. Consequently, the position of the stable circular orbit and its oscillation behavior differ significantly from predictions for Kerr BHs. Precession frequencies $\Omega_p$ Fig.~\ref{fig.8} indicates that increasing $a$ enhances the spacetime frame-dragging effect, leading to higher precession frequencies for particles moving near the charged QCBH. The parameter $b$ has a relatively small impact on spacetime geometry, and its effect on precession frequencies is negligible compared to the dominant effect of spin. The parameter $Q$ also alters the gravitational field of the QCBH, thus affecting orbital dynamics, especially in the strong-field region near the event horizon. Similarly, the Lense-Thirring precession frequency $\Omega_{LT}$ in Fig.~\ref{fig.9} increases with increasing $a$, as faster rotation enhances the frame-dragging effect. Conversely, increasing the $b$ slightly weakens the spinning frame-dragging effect. This decrease is much smaller than the enhancement effect of $a$. Furthermore, increasing $Q$ reduces the effective spinning effect. Overall, $a$ remains the primary factor determining precession frequency, while \textcolor{black}{the charge and quantum correction parameters to orbital dynamics are relatively weak. The potential astrophysical implications of these corrections should be assessed through future quantitative comparison with observational data.}
\\ In summary, our research shows that the influence of $b$ to $Q$ and $a$ on \textcolor{black}{charged} QCBH spacetime should have particle significance for particle dynamics and epicyclic oscillation frequencies.  
The theoretical corrections introduced by electric charged and quantum correction parameters to orbital dynamics are relatively weak. The potential astrophysical implications of these corrections should be assessed through future quantitative comparisons with observational data.
\appendix
\section{Radial frequency values}
\label{Radial}
The radial frequency values are given by
\begin{eqnarray} 
\Omega_1 &=& -\big(-3b^3\big({\mathcal{L}}-a{\mathcal{E}}\big)^2+b^2r^2-9\big(a^2+Q^2 \nonumber \\
&+& (-2+r)r\big){\mathcal{L}^2}+2a\big(9a^2+9Q^2+2r(-9 \nonumber \\ &+& 
4r)\big){\mathcal{L}}{\mathcal{E}}+ \big(-9a^4+6r^4 +a^2\big(-9Q^2+(18 \nonumber \\ &-& 7r)r\big)\big){\mathcal{E}^2} +r^6\big(a^4(-3Q^2+2r)-3\big(Q^2 \nonumber \\ &+& (-2+r)r\big)^3+a^2\big(-6Q^4+Q^2(2Q-9r)r \nonumber \\
&+& r^2(-12+r(6+r)))\big){\mathcal{L}^2}\big). \nonumber
\end{eqnarray} 
\begin{eqnarray} \nonumber
\Omega_2 &=& 3Q^2(a^2+Q)^2-2(a^2+Q^2)(a^2+9Q^2)r \nonumber\\ &+& 3\big(4a^2+3(4+a^2)Q^2+3Q^4\big). \nonumber
\end{eqnarray}
\begin{eqnarray}
\Omega_3 &=& -\big(6(4+a^2+6Q^2)r^3+2(16+5Q^2)r^4\nonumber \\&-&12r^5\big){\mathcal{L}}{\mathcal{E}}+\big(a^6(-3Q^2+2r)+a^4\big(-6Q^4\nonumber\\&+&Q^2(2Q-9r)r\big)+6(-2+r)r^2\big)+r^4(Q^4\nonumber\\&-&3Q^2r^2+2r^3). \nonumber
\end{eqnarray} 
\begin{eqnarray}
\Omega_4 &=& a^2\big(-3Q^6-9Q^4(-2+r)r-9Q^2-2\nonumber\\&+&r\big)^2r^2+2r^3\big(12+r(-14+3r)\big){\mathcal{E}}. \nonumber
\end{eqnarray}
\begin{eqnarray}
\Omega_5 &=& -\big(10a^4+9\big(Q^2+(-2+r)r\big)^2+a^2\big(19Q^2\nonumber\\&+&3r(-14+9r)\big)+r^2\big(36+r(-5Q\nonumber\\&+&21r)\big)\big).\nonumber
\end{eqnarray}
\begin{eqnarray}
\Omega_6 &=& -\big(10a^6+a^4\big(19Q^2+6r(-7+5r)\big) \nonumber\\ &+& r^4\big(-3Q^2+{2r}(-3+{5r})\big)+a^2\big(9{Q^4} \nonumber\\ &+& 12Q^{2r}(-3+{2r})+2{r^2}(18+r(32\nonumber\\&+&15r))\big)\big).\nonumber
\end{eqnarray}
\begin{eqnarray}
\Omega_7&=&r^4\big(a\big(b+(Q^2-2r)r^2\big){\mathcal{L}}+r^6{\mathcal{E}}+a^2-b\nonumber\\&+&r^2\big(Q^2+r(2+r)\big)\big){\mathcal{E}}^2.\nonumber
\end{eqnarray}
\vspace{-2.5mm}
\section{Vertical frequency values}
\label{vertical}
The vertical frequency values are given by
\begin{eqnarray} \nonumber
\Omega_8 &=& 4a^3\big(b+(Q^2-2r)r^2\big){\mathcal{L}}{\mathcal{E}}+a^4-2b\nonumber\\&+&r^2\big(-2Q^2+r(4+r)\big){\mathcal{E}^2}-a^2\big(b(2{\mathcal{L}^2}\nonumber\\&+&r^2{\mathcal{E}^2})\big). \nonumber
\end{eqnarray} 
\begin{eqnarray} \nonumber
\Omega_9 &=&r^6{\mathcal{E}+a^2\big(-b+r^2(-Q^2+r(2+r))\big){\mathcal{E}^2}}.\nonumber 
\end{eqnarray} 

\section*{Acknowledgements} 
The work was sponsored by the National Natural Science Foundation of China (Grant Numbers: 12235007, 12375003).

\section*{Data availability statement} 
This study does not involve any observational data. Calculations, analytical procedures, and figures supporting the results of this study are available from the corresponding author upon reasonable request.


\begin{thebibliography}{99}

\bibitem{Einstein1915}
A.~Einstein,
Die Feldgleichungen der Gravitation,
\textit{Sitzungsberichte der K{\"o}niglich Preu{\ss}ischen Akademie der Wissenschaften zu Berlin},
pp.~844--847, 1915.

\bibitem{Will2014}
C.~M.~Will,
The Confrontation Between General Relativity and Experiment,
\textit{Living Reviews in Relativity},
vol.~17, p.~4, 2014.

\bibitem{Abbott}
B.~P.~Abbott \textit{et al.},
Observation of Gravitational Waves from a Binary Black Hole Merger,
\textit{Physical Review Letters},
vol.~116, no.~6, p.~061102, 2016.

\bibitem{Schwarzchild1916}
K.~Schwarzschild,
{\"U}ber das Gravitationsfeld eines Massenpunktes nach der Einsteinschen Theorie,
\textit{Sitzungsberichte der K{\"o}niglich Preu{\ss}ischen Akademie der Wissenschaften zu Berlin},
pp.~189--196, 1916.

\bibitem{Riess1998}
A.~G.~Riess \textit{et al.},
Observational Evidence from Supernovae for an Accelerating Universe and a Cosmological Constant,
\textit{The Astronomical Journal},
vol.~116, no.~3, pp.~1009--1038, 1998

\bibitem{Plebanski and Krasinski2024}
J.~Pleba\'nski and A.~Krasi\'nski,
An Introduction to General Relativity and Cosmology,2nd~ed.,
\textit{Cambridge University Press}, 2024.

\bibitem{d'Inverno1992}
R.~d'Inverno,
Introducing Einstein's Relativity,
\textit{Clarendon Press}, Oxford, 1992.

\bibitem{Hobson2006}
M.~P.~Hobson, G.~E.~P.~Boxter, and A.~N.~Lasenby,
General Relativity: An Introduction for Physicists,
\textit{Cambridge University Press}, New York, 2006.

\bibitem{C.Liu2020}
C.~Liu, T.~Zhu, Q.~Wu, K.~Jusufi, M.~Jamil, M.~Azreg-A\"{\i}nou, and A.~Wang,
Shadow and Quasinormal Modes of a Rotating Loop Quantum Black Hole,
\textit{Physical Review D},
vol.~101, no.~8, p.~084001, 2020.

\bibitem{M.Afrin2023}
M.~Afrin, S.~Vagnozzi, and S.~G.~Ghosh,
Tests of Loop Quantum Gravity from the Event Horizon Telescope Results of Sgr A$^{*}$,
\textit{The Astrophysical Journal},
vol.~944, no.~2, p.~149, 2023.

\bibitem{Glampedakis}
K.~Glampedakis and G.~Pappas,
Is a Black Hole Shadow a Reliable Test of the No-Hair Theorem?,
\textit{Physical Review D},
vol.~107, no.~6, p.~064001, 2023.

\bibitem{Zhang}
J.~Lewandowski, Y.~Ma, J.~Yang, and C.~Zhang,
Quantum Oppenheimer-Snyder and Swiss Cheese Models,
\textit{Physical Review Letters},
vol.~130, no.~10, p.~101501, 2023.

\bibitem{Gong}
H.~Gong, S.~Li, D.~Zhang, G.~Fu, and J.-P.~Wu,
\textit{Physical Review D},
vol.~110, p.~044040, 2024.

\bibitem{Chen}
C.-Y.~Chen,
On the Possible Spacetime Structures of Rotating Loop Quantum Black Holes,
\textit{International Journal of Geometric Methods in Modern Physics},
vol.~19, no.~11, p.~2250176, 2022.

\bibitem{Ali}
H.~Ali, S.~U.~Islam, and S.~G.~Ghosh,
Shadows and Parameter Estimation of Rotating Quantum Corrected Black Holes and Constraints from EHT Observation of M87* and Sgr A$^{*}$,
\textit{Journal of High Energy Astrophysics},
vol.~47, p.~100367, 2025.

\bibitem{Vachher}
A.~Vachher and S.~G.~Ghosh,
Strong Gravitational Lensing by Rotating Quantum-Corrected Black Holes: Insights and Constraints from EHT Observations of M87$^{}$ and Sgr A$^{}$,
\textit{Journal of High Energy Astrophysics},
vol.~45, pp.~75--86, 2025.

\bibitem{Bardeen1972}
J.~M.~Bardeen, W.~H.~Press, and S.~A.~Teukolsky,
Rotating Black Holes: Locally Nonrotating Frames, Energy Extraction, and Scalar Synchrotron Radiation,
\textit{The Astrophysical Journal},
vol.~178, pp.~347--370, 1972.

\bibitem{Mashhoon1985}
B.~Mashhoon,
Stability of Charged Rotating Black Holes in the Eikonal Approximation,
\textit{Physical Review D},
vol.~31, no.~2, pp.~290--293, 1985.

\bibitem{Stuchlik2010}
Z.~Stuchl\'{\i}k and J.~Schee,
"Appearance of Keplerian Discs Orbiting Kerr Superspinars",
\textit{Classical and Quantum Gravity},
vol.~27, no.~21, p.~215017, 2010.

\bibitem{Kolos2017}
M.~Kolo\v{s}, A.~Tursunov, and Z.~Stuchl\'{\i}k,
Possible Signature of Magnetic Fields Related to Quasi-Periodic Oscillations Observed in Microquasars,
\textit{The European Physical Journal C},
vol.~77, p.~860, 2017.

\bibitem{Oteev2018}
T.~Oteev, M.~Kolo\v{s}, and Z.~Stuchl\'{\i}k,
Application of the Charged String Loop Oscillation Model to HF QPOs Observed in Microquasars,
\textit{The European Physical Journal C},
vol.~78, p.~764, 2018.

\bibitem{Ren2020}
S.~U.~Khan and J.~Ren,
Shadow Cast by a Rotating Charged Black Hole in Quintessential Dark Energy,
\textit{Physics of the Dark Universe},
vol.~30, p.~100644, 2020.

\bibitem{Zahid2021}
M.~Zahid, S.~U.~Khan, and J.~Ren,
"Shadow Cast and Center of Mass Energy in a Charged Rotating Black Hole with Perfect Fluid Dark Matter",
\textit{Chinese Journal of Physics},
vol.~72, pp.~575--586, 2021.

\bibitem{Bambi2017}
C.~Bambi,
Black Holes: A Laboratory for Testing Strong Gravity,
\textit{Springer}, Singapore, 2017.

\bibitem{Zhou2018}
M.~Zhou, Z.~Cao, A.~Abdikamalov, D.~Ayzenberg,
C.~Bambi, L.~Modesto, and S.~Nampalliwar,
Testing Conformal Gravity with the Supermassive Black Hole in 1H0707-495,
\textit{Physical Review D},
vol.~98, no.~2, p.~024007, 2018.

\bibitem{Kaluza1921}
T.~Kaluza,
Zum Unit\"{a}tsproblem der Physik,
\textit{Sitzungsberichte der Preussischen Akademie der Wissenschaften zu Berlin},
pp.~966--972, 1921.

\bibitem{Klein1926}
O.~Klein,
Quantentheorie und f\"unfdimensionale Relativit\"atstheorie,
\textit{Zeitschrift f\"ur Physik},
vol.~37, pp.~895--906, 1926.

\bibitem{Ortin2015}
T.~Ort\'{\i}n,
Gravity and Strings,2nd~ed.,
\textit{ Cambridge University Press}, Cambridge, UK, 2015.

\bibitem{Duff1986}
M.~J.~Duff, B.~E.~W.~Nilsson, and C.~N.~Pope,
Kaluza--Klein Supergravity,
\textit{Physics Reports},
vol.~130, nos.~1--2, pp.~1--142, 1986

\bibitem{Overduin1997}
J.~M.~Overduin and P.~S.~Wesson,
Kaluza--Klein Gravity,
\textit{Physics Reports},
vol.~283, nos.~5--6, pp.~303--378, 1997.

\bibitem{Gibbons1988}
G.~W.~Gibbons and K.-i.~Maeda,
Black Holes and Membranes in Higher-Dimensional Theories with Dilaton Fields,
\textit{Nuclear Physics B},
vol.~298, pp.~741--775, 1988.

\bibitem{Garfinkle1991}
D.~Garfinkle, G.~T.~Horowitz, and A.~Strominger,
Charged Black Holes in String Theory,
\textit{Physical Review D},
vol.~43, no.~10, pp.~3140--3143, 1991

\bibitem{MYAJ0} 
M. Yasir, X. Tiecheng, S. Chaudhary, and A. B. Jumah,  
\textit{J.High Energy Astrophys},
vol.~44, p.~356 2024.

\bibitem{MYAJ1}
M. Yasir, X. Tiecheng, and A. Jawad,
\textit{Eur. Phys. J. C} 
vol.~84, 946 2024.

\bibitem{MYAJ3} 
A. Jawad, M. Yasir, and H. Raza, 
\textit{Eur. Phys. J. C}
vol.~83, 882 2023.

\bibitem{Herdeiro2025}
C.~A.~R.~Herdeiro, E.~Radu, and E.~dos~Santos~Costa~Filho,
Charged, Rotating Black Holes in Einstein--Maxwell--Dilaton Theory,
\textit{Physical Review D},
2025.

\bibitem{Ashraf2025}
A.~Ashraf, A.~Bouzenada, S.~Maurya, F.~Atamurotov,
P.~Channuie, A.~Abd-Elmonem, and N.~Abdalla,
Rotating Black Holes in Einstein--Maxwell--Dilaton Theory:
From Weak to Strong Coupling,
\textit{Physics of the Dark Universe},
vol.~47, p.~101787, 2025.

\bibitem{Mustafa2025}
G.~Mustafa, P.~Channuie, F.~Javed, A.~Bouzenada,
S.~Maurya, A.~Cilli, and E.~G\"udekli,
Orbital Motion and Epicyclic Oscillations Around a Black Hole with Magnetic Charge,
\textit{Physics of the Dark Universe},
vol.~47, p.~101765, 2025.

\bibitem{Misner1973}
C.~W.~Misner, K.~S.~Thorne, and J.~A.~Wheeler,
Gravitation,
\textit{W.~H.~Freeman and Company}, San Francisco, 1973.

\bibitem{Hartle2003}
J.~B.~Hartle,
Gravity: An Introduction to Einstein's General Relativity,
\textit{Addison-Wesley}, San Francisco, 2003.

\bibitem{Witzany2019}
V.~Witzany,
A Null Geodesic View of the Kerr Spacetime,
\textit{Physical Review D},
vol.~100, no.~10, p.~104030, 2019.

\bibitem{Piovano2025}
G.~A.~Piovano, C.~Pantelidou, J.~Mac~Uilliam, and V.~Witzany,
Spinning Particles Near Kerr Black Holes: Orbits and Gravitational-Wave Fluxes Through the Hamilton--Jacobi Formalism,
\textit{Physical Review D},
vol.~111, no.~4, p.~044009, 2025.

\bibitem{Misner1973}
C.~W.~Misner, K.~S.~Thorne, and J.~A.~Wheeler,
Gravitation,
\textit{W.~H.~Freeman and Company}, San Francisco, 1973.

\bibitem{Chandrasekhar1983}
S.~Chandrasekhar,
The Mathematical Theory of Black Holes,
\textit{Oxford University Press}, New York, 1983

\bibitem{Wilkins1972}
D.~C.~Wilkins,
Bound Geodesics in the Kerr Metric,
\textit{Physical Review D},
vol.~5, no.~4, pp.~814--822, 1972.

\bibitem{Quevedo2011}
H.~Quevedo,
Geometrothermodynamics of Black Holes in Two Dimensions,
\textit{Physical Review D},
vol.~83, no.~2, p.~024012, 2011.

\bibitem{Kato1990}
S.~Kato, J.~Fukue, and S.~Mineshige,
Black-Hole Accretion Disks,
\textit{Kyoto University Press}, Kyoto, Japan, 1998.

\bibitem{Ryan1995}
F.~D.~Ryan,
Gravitational Waves from the Inspiral of a Compact Object into a Massive, Axisymmetric Body with Arbitrary Multipole Moments,
\textit{Physical Review D},
vol.~52, no.~10, pp.~5707--5718, 1995.

\bibitem{Levin2009}
J.~Levin and G.~Perez-Giz,
Homoclinic Orbits Around Spinning Black Holes. I. Exact Solution for the Kerr Separatrix,
\textit{Physical Review D},
vol.~79, no.~12, p.~124013, 2009.

\bibitem{Abramowicz2013}
M.~A.~Abramowicz and P.~C.~Fragile,
Foundations of Black Hole Accretion Disk Theory,
\textit{Living Reviews in Relativity},
vol.~16, p.~1, 2013.

\textcolor{black}{
\bibitem{Mustafa2024}
G. Mustafa, E. Demir, F. Javed, S. K. Maurya, E. Güdekli, 
S. Murodov, and F. Atamurotov,
Epicyclic oscillations and particle collision with trajectories around quantum corrected black holes,
\textit{Physics of the Dark Universe}, 
vol. 46, Art. no. 101708, 2024.
\bibitem{Zulqarnain2026}
R. M. Zulqarnain, O. Donmez, A. Bouzenada, I. Shakir, 
R. U. H. Muhammad, E. Güdekli, and S. Mardonov,
Testing particle dynamics, epicyclic oscillations around rotating 
Bumblebee Kerr--Newman-AdS black hole,
\textit{International Journal of Geometric Methods in Modern Physics},
Art. no. 2650222, 2026.
\bibitem{Ashraf2025}
A. Ashraf, F. Javed, S. K. Maurya, P. Channuie, A. Cilli, and E. Güdekli,
Testing of rotating Einstein--Yang--Mills--Higgs black hole through QPOs,
\textit{Physics of the Dark Universe},
vol. 48, Art. no. 101853, 2025.}

\bibitem{Yasir2024Barrow}
M. Yasir, X. Tiecheng, and A. Jawad,
Topological charges via Barrow entropy of black hole in metric-affine gravity,
\textit{Eur. Phys. J. C} \textbf{84}, 946 (2024).

\bibitem{Yasir2024Thermal}
M. Yasir, X. Tiecheng, F. Javed, and G. Mustafa,
Thermal analysis and Joule--Thomson expansion of black hole exhibiting metric-affine gravity,
\textit{Chin. Phys. C} \textbf{48}, 015103 (2024).

\bibitem{Chaudhary2022Thermodynamic}
S. Chaudhary, A. Jawad, and M. Yasir,
Thermodynamic geometry and Joule--Thomson expansion of black holes in modified theories of gravity,
\textit{Phys. Rev. D} \textbf{105}, 024032 (2022).

\bibitem{Yasir2025Particle}
M. Yasir, F. Mushtaq, X. Tiecheng, and F. Javed,
Investigating the effects of particle motion and gravitational lensing of black hole in string inspired Euler--Heisenberg theory,
\textit{Phys. Dark Univ.} \textbf{48}, 101838 (2025).

\bibitem{Mushtaq2025Imprints}
F. Mushtaq, X. Tiecheng, M. Yasir, F. Javed, M. Alosaimi, and R. M. Zulqarnain,
Imprints of magnetic charge on the particle orbits, weak gravitational lensing and black hole shadows,
\textit{Phys. Dark Univ.} \textbf{50}, 102109 (2025).
\end{thebibliography}
\end{document}